\documentclass[aps,prb,twocolumn,superscriptaddress,amsmath,amssymb,amsfonts,showpacs]{revtex4}

\usepackage{graphicx}
\usepackage{bm}
\usepackage{bbold}
\usepackage{color}
\usepackage[T1]{fontenc}

\newcommand{\bea}{\begin{eqnarray}}

\newcommand{\eea}{\end{eqnarray}}

\newcommand{\beq}{\begin{equation}}

\newcommand{\eeq}{\end{equation}}

\newenvironment{frcseries}{\fontfamily{frc}\selectfont}{}
\newcommand{\textfrc}[1]{{\frcseries#1}}

\newcommand{\figref}[1]{Figure~\ref{#1}}

\newcommand{\secref}[1]{Section~\ref{#1}}
\newcommand{\subsecref}[1]{Subsection~\ref{#1}}
\newcommand{\appref}[1]{Appendix~\ref{#1}}
\newcommand{\Eqref}[1]{Eq.~(\ref{#1})}
\newcommand{\mathfrc}[1]{\text{\scriptsize{\textfrc{#1}}}}

\newlength{\textwidthm}

\begin{document}

\def \tr{{\mbox{tr~}}}
\def \ra{{\rightarrow}}
\def \ua{{\uparrow}}
\def \da{{\downarrow}}
\def \ba{\begin{array}}
\def \ea{\end{array}}
\def \nn{\nonumber}
\def \l{\left}
\def \rg{\right}
\def \half{{\frac{1}{2}}}
\def \etal{{\it {et al}}}
\def \cH{{\cal{H}}}
\def \cE{{\cal{E}}}
\def \cK{{\cal{K}}}
\def \cM{{\cal{M}}}
\def \cN{{\cal{N}}}
\def \cQ{{\cal Q}}
\def \cI{{\cal I}}
\def \cV{{\cal V}}
\def \cD{{\cal D}}
\def \cP{{\cal P}}
\def \cF{{\cal F}}
\def \cZ{{\cal Z}}
\def \cC{{\cal C}}
\def \cO{{\cal O}}
\def \cS{{\cal S}}
\def \bS{{\bf S}}
\def \bI{{\bf I}}
\def \bL{{\bf L}}
\def \bG{{\bf G}}
\def \bQ{{\bf Q}}
\def \bK{{\bf K}}
\def \bR{{\bf R}}
\def \br{{\bf r}}
\def \bu{{\bf u}}
\def \bp{{\bf p}}
\def \bq{{\bf q}}
\def \bk{{\bf k}}
\def \bz{{\bf z}}
\def \bx{{\bf x}}
\def \bpsi{{\overline{\psi}}}
\def \tA{{\tilde{A}}}
\def \tC{{\tilde{C}}}
\def \tu{{\tilde{u}}}
\def \K{{\kappa}}
\def \W{{\Omega}}
\def \tW{{\tilde{\Omega}}}
\def \lam{{\lambda}}
\def \L{{\Lambda}}
\def \a{{\alpha}}
\def \t{{\theta}}
\def \T{{\Theta}}
\def \b{{\beta}}
\def \g{{\gamma}}
\def \D{{\Delta}}
\def \d{{\delta}}
\def \w{{\omega}}
\def \s{{\sigma}}
\def \f{{\phi}}
\def \F{{\Phi}}
\def \x{{\chi}}
\def \e{{\epsilon}}
\def \h{{\eta}}
\def \G{{\Gamma}}
\def \z{{\zeta}}
\def \hn{{\overline{n}}}
\def \nd{{^{\vphantom{\dagger}}}}
\def \yd{{^\dagger}}

\def \gs{{\gamma_S}}
\def \ml{{\mathfrc{l}}}
\def \Etd{{\text{E}_{\text{\tiny{2D}}}}}
\def \ETotBD{{E^{\text{\tiny{Tot}}}_{\text{\tiny{BD}}}}}
\def \ETotSD{{E^{\text{\tiny{Tot}}}_{\text{\tiny{SD}}}}}

\title{ Pinning of a two-dimensional membrane on top of a patterned substrate:
   The case of graphene.}

\author{S.~Viola~Kusminskiy}\affiliation{Department of Physics, Boston University, 590 Commonwealth
  Ave., Boston, MA 02215.\footnote{Current address: Dahlem Center for Complex Quantum Systems
Freie Universitaet
Arnimallee 14, 14195, Berlin Germany. }}
\author{D.K Campbell} \affiliation{Department of Physics, Boston University, 590 Commonwealth
  Ave., Boston, MA 02215.}
\author{A. H. {Castro Neto}}\affiliation{Department of Physics, Boston University, 590 Commonwealth
  Ave., Boston, MA 02215.}
\author{F. Guinea}\affiliation{Instituto de Ciencia de Materiales de Madrid, CSIC, Sor Juana Ines de la Cruz 3, E28049, Madrid Spain.}

\date{\today}

\begin{abstract}
We study the pinning of a two-dimensional membrane to a patterned substrate within elastic theory both in the bending rigidity and in the strain dominated regimes. We find that both the in-plane strains and the bending rigidity can lead to depinning. We show from energetic arguments that the system experiences a first order phase transition between the attached configuration to a partially detached one when the relevant parameters of the substrate are varied, and we construct a qualitative phase diagram. Our results are confirmed through analytical solutions for some simple geometries of the substrate's profile. We apply our model to the case of graphene on top of a SiO2 substrate and show that typical orders of magnitude for corrugations imply graphene will be partially detached from the substrate.
\end{abstract}


\maketitle
\section{Introduction}\label{sec:Intro}
Until recently, the study of two dimensional (2D) membranes was developed mainly for its theoretical interest and its applications to biological systems which could be well approximated by the 2D membrane model, as well as soft matter systems ~\cite{NelsonBook}. Nowadays, however, with the experimental discovery of graphene ~\cite{Novoselov04,Novoselov05,Kim05,Geim_review} (a two dimensional graphite sheet), we have in our hands the opportunity of studying a truly 2D membrane. It has been proven that the membrane aspect of graphene, and in particular the presence or not of a substrate, plays an essential role to characterize its behavior ~\cite{Morozov06,Morozov08,Fuhrer08}. Graphene presents intrinsic ripples ~\cite{Meyer07}, inherent to its 2D nature, which can interact with the propagating electrons and affect transport properties ~\cite{AntonioRMP}. In most experimental settings up to date, though, graphene is deposited on top of a substrate, either purposely patterned or presenting random disorder. A relevant question then is to determine how the spatial structure of the substrate affects that of graphene. This kind of study also opens the possibility of controlling the properties of graphene by patterning appropriately the substrate. Experiments have shown that the morphology of a graphene membrane on top of a substrate is largely determined by the substrate's profile ~\cite{Fuhrer07,Stolyarova07,Morgenstern09,Scharfenberg10}, as opposed to suspended graphene. The attachment of graphene to a
corrugated surface leads to the bending and stretching of the
graphene layer, so that the
depinning of the layer may become energetically favorable. For device-construction, as well as for the interpretation of experimental data, it is important to know if the graphene sheet is completely pinned to the substrate or if there are regions for which depinning occurs and graphene is suspended.

Given its experimental relevance, in this work we address the problem of determining which is the stable configuration of a membrane on top of a substrate which presents either depressions or protrusions.  Although we will treat the problem in the context of graphene physics, our results are general. We analyze this problem from a general field theory framework, in which we show the possibility of a phase transition between a pinned configuration to a partially depinned one, where relevant parameters of the patterned substrate act as control parameters. We turn then to analyze some simple substrate geometries which allow for analytical solutions. We will show that these examples quantitatively  confirm our phenomenological, qualitative model. Our work is the first in the graphene literature that takes into account the effect of in-plane strains for detachment of the membrane. We show that there is a length scale for the substrate's pattern beyond which the in-plane strains are dominant and can lead to depinning. This length scale marks the crossover from a regime in which the bending rigidity of the membrane is dominant energetically.

In what follows we will analyze the depinning of a membrane from the substrate for the two different limiting regimes mentioned above ~\footnote{Previous works have studied this problem for graphene restricted to the bending rigidity dominated regime and for some particular patterns of the substrate ~\cite{Pierre-Louis08,TengLi10a,TengLi10b}. A comprehensive treatment of the problem of adhesion of a membrane to a substrate, in the BD regime, can be found in Ref. \onlinecite{Andelman99,Andelman01}.}. Firstly, in \secref{sec:Model} we introduce the model for the free energy of a membrane on top of a substrate and by means of scaling arguments we establish the possibility of a phase transition for the system between two possible stable equilibrium configurations: the membrane being completely attached to the substrate, or otherwise it being partially detached. From this we are able to construct a qualitative phase diagram for the system. In the following sections we proceed to a quantitative analysis for a given geometry of the substrate profile. We consider a substrate with a Gaussian depression or protuberance and we obtain analytical solutions for the two limiting regimes, the bending rigidity dominated regime in \secref{sec:FlexModes} and the strain dominated regime in \secref{sec:InPlaneModes}. In both cases, we show that the system presents a first order phase transition from pinned-to-depinned as the ratio of width to height of the substrate's profile is varied. A discussion and possible experimental consequences are presented in \secref{sec:Disc}.

\section{Model and qualitative phase diagram}
\label{sec:Model}
\begin{figure}
 \centering
 \includegraphics[width=8cm]{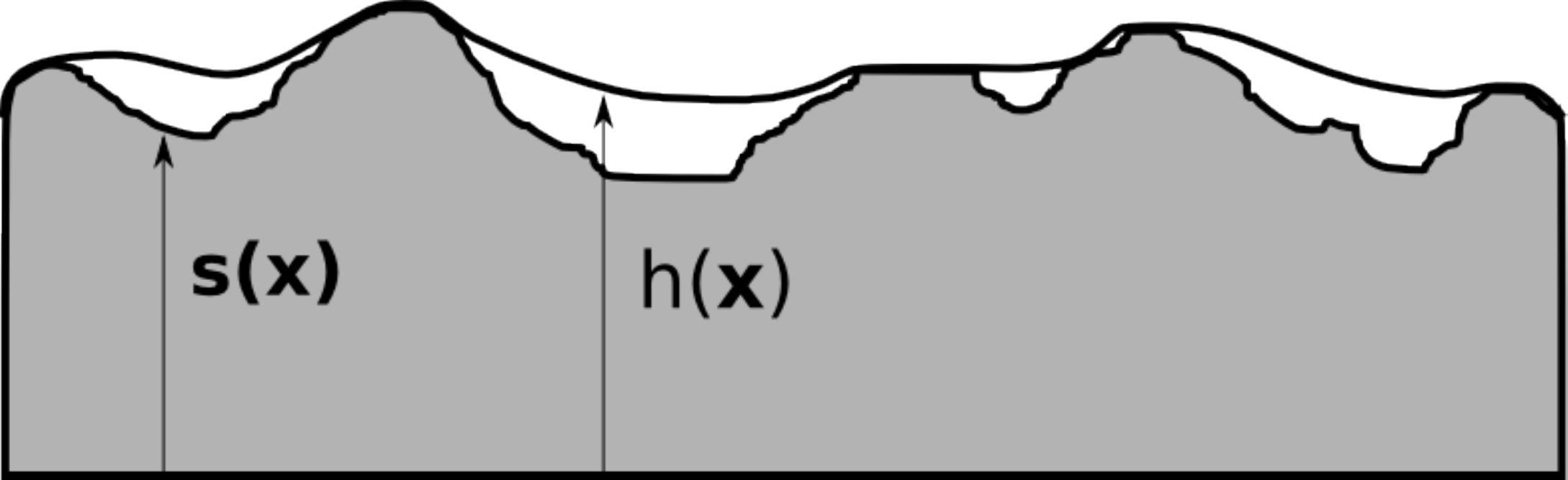}
 \caption{Pictorial representation of a membrane on top of a random substrate, partially conforming to the substrate.The height of the membrane is represented by a field $h(\bx)$ while the top surface of the substrate is represented by a field $s(\bx)$, as discussed in the main text.}
 \label{fig:GrapheneOnSubs}
\end{figure}

We consider a tethered membrane which lies on top of a substrate. We use the {\it de Monge} parametrization ~\cite{LubenskyBook}, by which the membrane is parametrized by $(\bx,h(\bx))$, where $h$ is the height with respect to some reference plane and $\bx=(x,y)$ are the in-plane coordinates. In the same way, the profile of the substrate is represented by $(\bx,s(\bx))$, as shown schematically in \figref{fig:GrapheneOnSubs}.

We assume as a first approximation that the membrane couples to the substrate through its out-of-plane modes (also denominated flexural modes), via a contact force characterized by a surface tension $\gs$. Previous works that study the attachment of a membrane to a substrate, have used the so called Deryagin approximation, which approximates the interaction potential between the membrane and the substrate as a harmonic potential ~\cite{Andelman99}. However this approximation results in a strongly confining potential. In our case we are interested in studying the stability of the pinned configuration and the possibility of detachment, and therefore a contact force approximation is more appropriate. Moreover, the interaction between graphene and a substrate has been studied in Ref. \onlinecite{Sabio08} and it has been shown that the attractive interaction force decays as the inverse distance to a power that depends on the type of interaction (a power of 2 in the case of undoped SiO$_2$). In that work it was also shown that the coupling strength decays roughly four orders of magnitude when the graphene sheet is not pinned to the substrate. These considerations justify the use of a contact force which is finite when graphene is conforming to the substrate and zero otherwise. This is, of course, an idealization of our model since we are disregarding the equilibrium distance between the substrate and the membrane, which for graphene on a SiO$_2$ substrate is of the order of $5$ \AA ~\cite{Fuhrer07}. The free energy for the membrane on top of the substrate within this approximation is given by
\beq\label{FreeEn}
\begin{split}
\cF[\bu,h,s]&=\half \int d^2x\l[ \K \l(\nabla^2h(\bx)\rg)^2+2\mu \tilde{u}_{ij}(\bx)^2+\lam \tilde{u}_{ii}(\bx)^2\rg]\\
&-\frac{1}{2}\gs\int_\cS d^2x
\end{split}
\eeq
where $\mu$ and $\lam$ are the Lam\'e coefficients and $\K$ is the bending rigidity of the membrane, and $\cS$ is the surface of contact between the membrane and the substrate. Throughout this paper we will use the accepted values of the elastic and bending parameters for graphene at room temperature. The bending rigidity is given by $\K \approx 1$ eV ~\cite{Fasolino07}, and the Lam{\'e} coefficients are given by $\mu \approx 10$ eV \AA$^{-2}$ and $\lam \approx 2$ eV \AA$^{-2}$ ~\cite{Fasolino09}. We take the value of the coupling constant strength as $\gs=2$ meV$\text{\r{A}}^{-2}$, corresponding to the maximum estimated pinning strength for graphene on a SiO$_2$ substrate ~\cite{Sabio08} ~\footnote{The value of $\gs\approx 2$meV \AA$^{-2}$ is actually an upper limit for $\gs$, attained for when there is a layer of water between graphene and the substrate ~\cite{Sabio08}. Also, repulsive forces not considered in Ref. \onlinecite{Sabio08} would effectively reduce the value of $\gs$. }.The functional dependence of $\cF[\bu,h,s]$ on the substrate's profile field $s(\bx)$ is given implicitly through the contact term, being $h(\bx)\equiv s(\bx)$ when the membrane is attached to the substrate.
$\l(\nabla^2h(\bx)\rg)^2$ is the local mean curvature of the membrane and the {\it local} intrinsic curvature is encoded in the strain tensor ~\footnote{Note that although the {\it total} Gaussian curvature of a nearly flat 2D membrane is zero, the quartic interaction term generated by integrating out the in-plane phonon modes can be seen as a long range interaction between the {\it local} Gaussian curvatures at different points in the membrane (see Ref. \onlinecite{LubenskyBook}). This long range interaction is the responsible for the stability of a low temperature flat phase in 2D membranes (see Refs. \onlinecite{NelsonBook,Peliti87}).}: $\tilde{u}_{ij}=\half\l(\partial_i\bu_j+\partial_j\bu_i+\partial_ih\partial_jh\rg)$, with $\bu(\bx)$ the in-plane phonon modes and the $i$, $j=1,2$ index the two components of the field. Since the action is quadratic in these modes, they can be integrated out ~\cite{Peliti87} to obtain an effective free energy $e^{-\cF_{eff}[h,s]}=\int \cD\bu\, e^{-\cF[\bu,h,s]}$
with
\beq\label{FreeEnEffTot}\begin{split}
\cF_{eff}[h,s]&=-\frac{\gs}{2}\int_\cS d^2x + \frac{\K}{2}\int d^2x \l(\nabla^2h(\bx)\rg)^2\\
&+ \frac{\Etd}{8}\int d^2x\l( P_{ij}^T\partial_ih(\bx)\partial_jh(\bx)\rg)^2 \, ,
\end{split}\eeq
being $P_{ij}^T=\d_{ij}-\frac{\partial_i\partial_j}{\nabla^2}$ the transverse projector and where we have used the expression for the Young modulus in 2D, $\Etd=\frac{4\mu(\mu+\lam)}{2\mu+\lam}$. We are interested in analyzing the possible detachment of the graphene sheet from the substrate, and in particular, to find the configuration which is energetically favorable. The general procedure would be to minimize the free energy \Eqref{FreeEnEffTot} given a profile of the substrate to find the stable solution. However, the non-linearity of \Eqref{FreeEnEffTot} makes this program impossible to follow analytically, even for the most simple geometries. We are then obliged to make use of approximations if we are to make any analytical progress. It is usually assumed that the in-plane stresses are small and therefore their contribution, encapsulated in the quartic order term of the effective energy $\cF_{eff}[h,s]$, can be neglected. However this is true only if the height fluctuations are not too big, as we proceed to show. If we consider a substrate of average height
fluctuations $S$ over a length scale $L$, from \Eqref{FreeEn} we see that the bending energy of a membrane attached to this substrate scales as
\beq\label{EKscaling}
E_K\sim \K\int d^2x \l(\nabla^2h(\bx)\rg)^2\sim\frac{\kappa}{L^{2}} S^2\,,
\eeq
where we have used that $\nabla\sim L^{-1}$ and the area $\int d^2x\sim L^2$.
On the other hand, by similar arguments (note that $\tilde{u}_{ij}\sim S^2/L^2$), the elastic energy due to in-plane strains is given roughly by
\beq\label{Eelscaling}
E_{el} \sim \frac{\Etd}{L^{2}} S^4.
\eeq
Therefore the elastic energy due to in-plane strains
is the main contribution to the total energy of the membrane if $\Etd \bar{s}^2
\gg \kappa$.
This analysis is valid except for quasi one dimensional (1D) geometries, where the height
profile of the substrate is constant along one direction. For this case it is easy to show that the in-plane strains are completely screened by the height fluctuations and hence the in-plane stresses are zero, and the only contribution to the elastic
energy is due to the bending rigidity.

With the previous analysis we have then arrived to a length scale
\beq\label{lenghtscale}
\mathfrc{l}=\sqrt{\frac{\K}{\Etd}}
\eeq
that determines a crossover from a {\it bending rigidity} dominated regime (BD regime) for $S<\ml$, to a {\it strain} dominated regime (SD regime) for $S>\ml$. With the values for the elastic parameters of graphene given above, $\ml\approx1$ \AA ~\footnote{Compare with the temperature dependent crossover length scale obtained by the renormalization of the bending rigidity due to the in-plane modes, see Refs. \onlinecite{NelsonBook,Safran07}.}. Note that this scale is of the order of magnitude of the lattice spacing and in principle this would imply that the BD regime for graphene is greatly suppressed ~\cite{Atalaya08}. However recent atomistic simulations have shown that thermal height fluctuations of this magnitude are possible ~\cite{Los09}. Moreover, the study presented in Ref. \onlinecite{Los09} shows that the continuum model can still be applied in this limit. This scale can thus be realized in graphene~\cite{Fuhrer07,Stolyarova07} and therefore the crossover is of experimental relevance.

We can now study the two limiting regimes separately. For the BD regime, the free energy for the membrane can be approximated by:
\beq\label{KFreeEn}
\cF_{eff}[h,s]\approx-\frac{\gs}{2}\int_\cS d^2x + \frac{\K}{2}\int d^2x \l(\nabla^2h(\bx)\rg)^2\, .
\eeq
To solve for the equilibrium configuration, we look for the saddle point solutions of equation \eqref{KFreeEn} with a partially detached membrane and study their stability. Minimizing with respect to the height $h(\bx)$ yields the bi-harmonic equation within the detached region:
\beq\label{biharm}
\l(\nabla^2\rg)^2 h(\bx)=0
\eeq
to be solved together with the appropriate boundary conditions, while $h(\bx)\equiv s(\bx)$ in the pinned region. The boundary conditions have to be imposed at the boundary of the surface $\cS$, that is the curve at which the membrane starts to detach from the substrate. If we parametrize this closed curve by $\bx^*\equiv \partial\cS$, the boundary conditions are given by:
\beq
\label{BC1}
h(\bx^*)=s(\bx^*)
\eeq
\beq
\label{BC2}
\nabla h(\bx^*)=\nabla s(\bx^*)\,.
\eeq
The curve $\bx^*$ itself is unknown, and can be determined by an extra boundary condition which implies a discontinuity in the second derivatives due to the surface tension force at the curve of detachment $\bx^*$ ~\cite{LandauElasticityBook}:
\beq
\label{BC3}
\gs=\K\l[\nabla^2 h(\bx^*)-\nabla^2 s(\bx^*)\rg]^2
\eeq
Alternatively, it is equivalent to find the extrema of the free energy \Eqref{KFreeEn} as a function of $\bx^*$. For a given, arbitrary profile of the substrate $s(\bx)$, it is to be expected that the free energy will have many extrema, corresponding to unstable and metastable configurations. The curve $\bx^*_0$ corresponding to a global minimum will give the stable equilibrium configuration, if this is the null curve then the stable configuration is the totally pinned membrane. We see then that the curve $\bx^*$ emerges as a natural order parameter of the problem, between two possible states of the system: a null curve $\bx^*_0\equiv0$ corresponding to a membrane which is completely attached to the substrate, and a finite value of the function $\bx^*_0$ which gives a partially detached membrane. A scalar order parameter can be obtained, for example, by taking the total length of the curve $\l|\bx^*\rg|$ ~\footnote{Note that in this general context, the curve $\bx^*$ can be disconnex.}. Our analytical results for the particular geometries studied, to be developed in the following sections, show that the pinned configuration is always at least a metastable minimum and hence the pinned-to-depinned transition is always of first order. We can argue this has to be true in general for smoothly corrugated substrates as follows. If we consider a small deviation of the system from the totally attached configuration $\l|\bx^*_0\rg|=0$, described by a small detachment curve $\l|\d\bx^*\rg|$, the energy cost due to depinning is proportional to the minimal area enclosed by the curve, $\sim\l|\d\bx^*\rg|^2$. On the other hand, the smoothness of the substrate implies that, for small enough $\l|\d\bx^*\rg|$, the area delimited by this curve is locally flat and hence the gain in energy due to the relaxation of bending and stretching of the membrane is negligible. Hence the pinned configuration is always a local minimum of the energy and the phase transition to a partially detached configuration is of first order, due to the development of new metastable states with the variation of the control parameters. It is safe to assume that, for fixed external conditions, these control parameters will be related to the characteristic width and height of the substrate's corrugations. To simplify the analysis, we can consider the problem of a single depression or protuberance in the substrate. Intuitively it is to be expected that the stability of the pinned configuration, given a coupling strength $\gs$ and bending rigidity $\K$, will depend on the aspect ratio of the substrate's profile. A simple energetic argument gives an estimate for this threshold. The interaction energy between the graphene layer and
the substrate in a region of area $L^2$ is
\beq\label{Epinscaling}
E_{pin}\sim\gs L^2\,,
\eeq
while, as we saw previously, the
bending energy cost of height corrugations of scale $S$ is given by \Eqref{EKscaling}. The change between the regime where the pinning energy is
dominant and the layer is attached to the substrate, to the regime where the cost in bending energy leads to the detachment of
the layer, is governed by the ratio
\beq\label{EpinToEK}
\frac{E_{pin}}{E_{K}}\sim\frac{\gs}{\K}\frac{L^4}{S^2}\,.
\eeq
The membrane will prefer to attach to the substrate in the limit $\frac{E_{pin}}{E_{K}}>1$, which translates into a condition for the substrate profile:
\beq\label{thresholdK}
\frac{S}{L^2}<\sqrt{\frac{\gs}{\K}}\,,
\eeq
indicating that pinning is favored for shallower depressions.

Within the bending rigidity approximation, detachment can occur due to the high bending energy cost that competes with the energy gain due to pinning. In the opposite regime, $S>\ml$, the in-plane stresses are dominant and we should consider the possible detachment due to these modes. For this case the free energy \eqref{FreeEn} can be approximated by:
\beq\label{IPFreeEn}
\cF\approx-\frac{\gs}{2}\int_\cS d^2x+\half \int d^2x \l[2\mu \tilde{u}_{ij}(\bx)^2+\lam \tilde{u}_{ii}(\bx)^2\rg]\, .
\eeq
In this limit, the approximate free energy given by \Eqref{IPFreeEn} still contains the non-linear coupling between the in-plane and out-of-plane modes and hence further approximations are necessary for obtaining analytical results ~\footnote{Note that we have to retain the non-linear term in the strain tensor since we are in the limit of large out-of-plane fluctuations.}. We will introduce these approximations in \secref{sec:InPlaneModes} when we solve the system for a particular geometry of the substrate. For now however, we can perform a scaling analysis similar to the one we did for the bending energy to determine a threshold energy for the pinned-to-depinned transition due to in-plane strains, depending on the aspect ratio of the perturbation in the substrate. In this case the transition is controlled by the ratio of pinning energy to elastic energy:
\beq
\frac{E_{pin}}{E_{el}}\sim\frac{\gs}{\Etd}\frac{L^4}{S^4}\,,
\eeq
where we have used Eqs. \eqref{Eelscaling} and \eqref{Epinscaling}. As in the previous case, we can argue that the membrane will favor the pinned configuration when $\frac{E_{pin}}{E_{el}}>1$, which gives us the condition:
\beq\label{thresholdEl}
\frac{S}{L}<\l(\frac{\gs}{\Etd}\rg)^{1/4}\,,
\eeq
again consistent with the intuitive picture that shallower depressions should favor pinning. The possible equilibrium solutions for the curve of detachment $\l|\bx^*\rg|$ are in this case given by the extrema of the free energy \Eqref{IPFreeEn}, as in the BD regime, a globally stable solution with $\l|\bx^*\rg|=0$ corresponds to the completely pinned configuration.

\begin{figure}
 \centering
 \includegraphics[width=8cm]{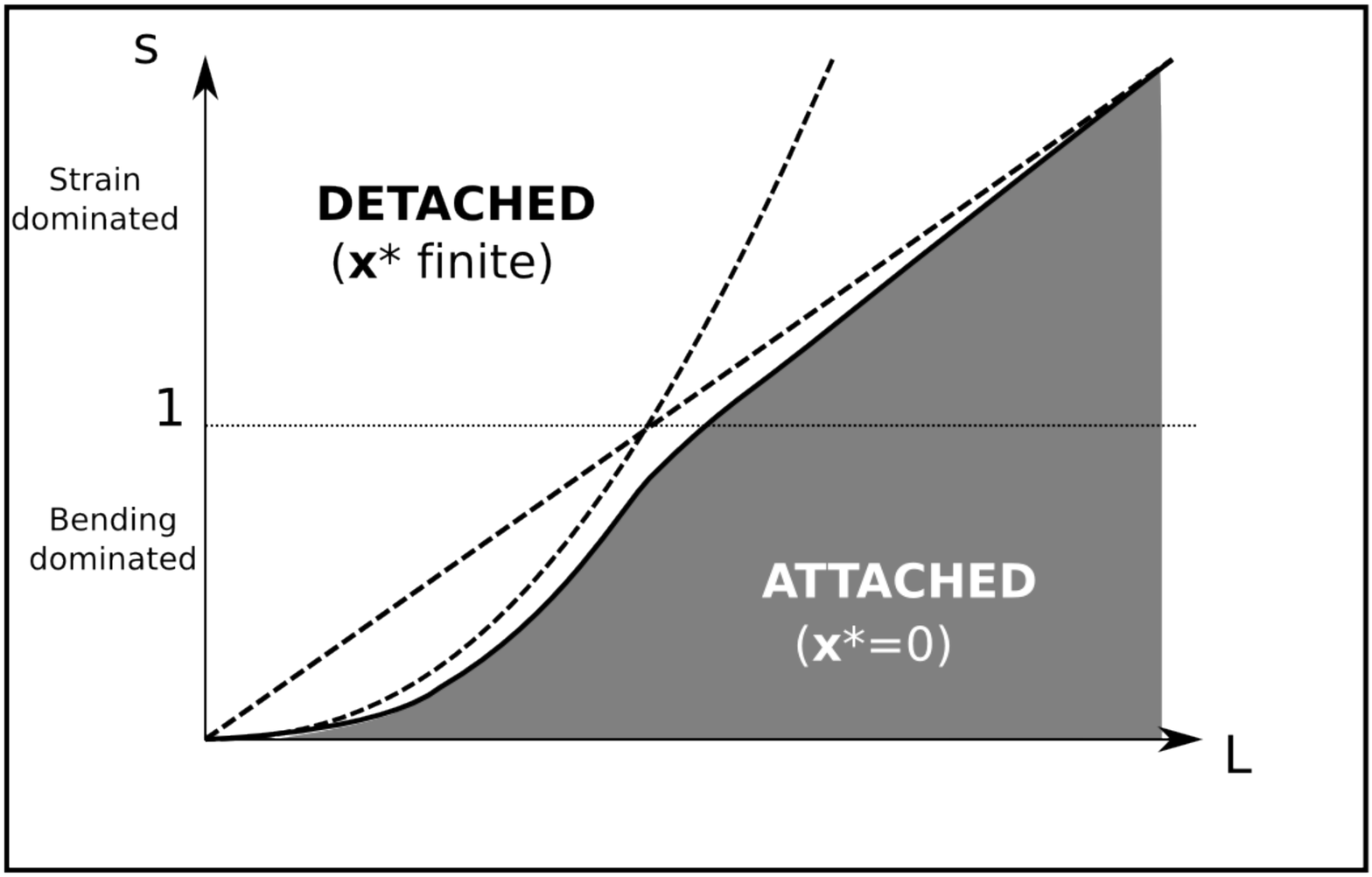}
 \caption{Qualitative phase diagram for a membrane on top of a patterned substrate of characteristic width $L$ and height $S$, in units of a the length scale $\ml$. The dashed lines correspond to the critical lines given by \Eqref{critlines}, the solid line is an estimated interpolation. Note that this phase diagram is not valid for 1D geometries, as discussed in the text. }
 \label{fig:phasediag}
\end{figure}

Equations \eqref{thresholdK} and \eqref{thresholdEl} define two lines of critical values given by $S_c\equiv S(L_c)$, which mark the transition from pinned-to-depinned in the parameter space of height and width of the substrate's profile. Note that while in the BD regime the dependence of $S_c$ on the critical width $L_c$ is quadratic (see \Eqref{thresholdK}), in the SD regime this dependence is linear. In the intermediate region hence it is to be expected a crossover between the two critical lines. These considerations allow us to construct a qualitative  phase diagram. For this it is useful to consider the dimensionless quantities $S\ra S/\ml$, $L\ra L/\ml$ that give the height and width of the substrate's profile in units of the length scale $\ml$ of the BD to SD regime crossover defined in \Eqref{lenghtscale}. We can then write the critical lines as:
\beq\label{critlines}
\begin{split}
 S_c&=\ml\sqrt\frac{\gs}{\K}L_c^2\quad  S_c\ll1\\
 S_c&=\l(\frac{\gs}{\Etd}\rg)^{1/4} L_c\quad  S_c\gg1
\end{split}
\eeq
The qualitative phase diagram is shown in \figref{fig:phasediag}.

Near the critical line the free energy \Eqref{FreeEn} can be written as a Landau functional of the order parameter $\l|\bx\rg|$:
\beq\label{LandauFreeEn}
\cF[\l|\bx\rg|]=A_2\l|\bx^*\rg|^2+A_3\l|\bx^*\rg|^3+...+A_n\l|\bx^*\rg|^n\, ,
\eeq
where $A_2$ is a positive constant (in accordance with $\l|\bx^*_0\rg|=0$ being always a local minimum) and the coefficients $A_i$, $i=3\,,...n$ are functions of the control parameters $S$, $L$. As usual, the expansion is cut at order $n>3$, being $A_n$ the first non-negative coefficient. The powers appearing in the expansion are dictated by the symmetry of the system, for cylindrically symmetric geometries only even powers are allowed. In the following sections we will re-obtain these results in an analytical fashion for certain simple geometries of the substrate.

\section{Detachment due to out-of-plane modes for radial symmetry}
\label{sec:FlexModes}

\begin{figure}
 \centering
 \includegraphics[width=8cm]{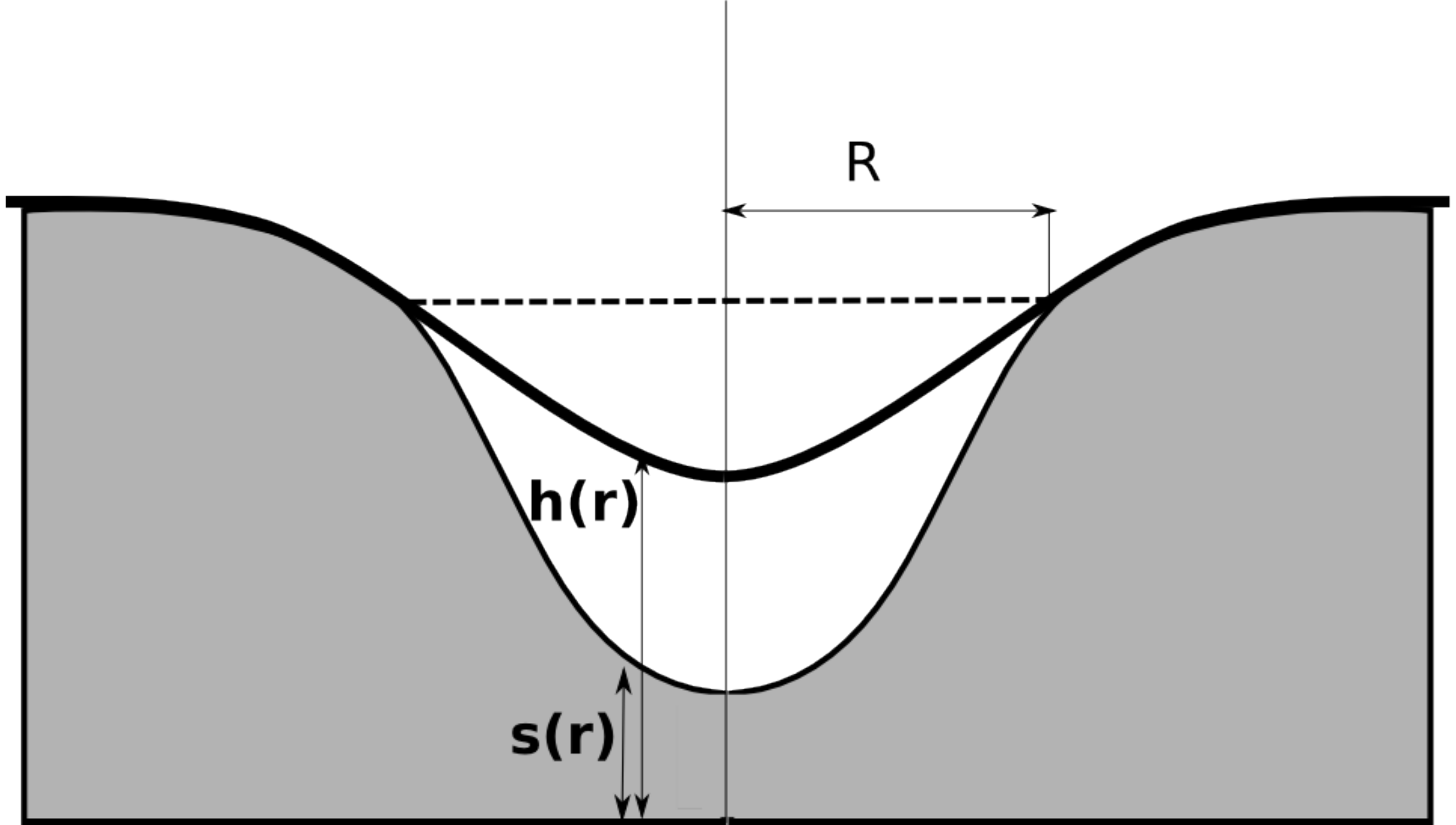}
 \caption{Membrane on top of a substrate with a depression. The figure is axially symmetric with respect to the vertical axis through the center of the substrate's depression. R indicates the radius of detachment. The dashed line represents the approximation used in \secref{sec:InPlaneModes} for when the in-plane modes are taken into account.}
 \label{fig:GrapheneOnGaussHole}
\end{figure}

As we stated in \secref{sec:Model}, for 1D geometries the solution obtained by only considering the bending rigidity term in the elastic free energy, is exact. For three dimensional (3D) geometries this is an approximation that works well for small height fluctuations of the substrate. Our aim in this section is to obtain analytical results in this limit to obtain a qualitative understanding of the depinning process. Analytical results can be obtained for certain simple geometries, we will restrict our analysis to cylindrically symmetric cases. We consider first a substrate with an axially symmetric {\it depression} $s(r)$ as shown in \figref{fig:GrapheneOnGaussHole}.
The bi-harmonic equation \eqref{biharm} in cylindrical coordinates, assuming a rotational invariant case, is given by:
\beq\label{biharmcyl}
\l(\frac{1}{r}\partial_r+\partial_r^2\rg)^2h(r)=0
\eeq
which has the following general solution:
\beq\label{GenSolh}
h(r)=H_0+H_1\log{r}+\frac{H_2}{2}r^2+H_3r^2\log{r}
\eeq
For a depression, if we assume that the membrane detaches from the substrate homogeneously at a circumference of radius $R$ (to be determined), the solution \Eqref{GenSolh} is valid for $0\le r\le R$ and hence it has to be regular at the origin, $H_1=H_3=0$. The radius $r=R$ gives the parametrization of the curve of detachment $\bx^*$ introduced in \secref{sec:Model}. The boundary conditions \Eqref{BC1}-\Eqref{BC3} take the form:
\beq\label{radBC}
\begin{split}
h(R)&=s(R)\\
h'(R)&=s'(R)\\
\frac{h'(R)}{R}+h''(R)&= \frac{s'(R)}{R}+s''(R)\pm\sqrt{\frac{\gs}{\K}}\, .
\end{split}
\eeq
Applying the boundary conditions \Eqref{radBC} over the general solution \Eqref{GenSolh} we obtain:
\beq
\begin{split}\label{radSol}
h(r)&=s(R)-\frac{s'(R)}{2}R +\frac{s'(R)}{2R}r^2\\
\frac{s'(R)}{R}&=s''(R)\pm\sqrt{\frac{\gs}{\K}}\,.
\end{split}
\eeq
The second equation determines the radius of detachment, but also imposes a condition over the substrate profile for a non-trivial solution to exist (note that there is always a solution with $R=0$). As we pointed out in \secref{sec:Model}, the radius of detachment $R$ corresponds to extrema of the total energy of the membrane, which within the present approximation consists of
\beq\label{ETotBDDef}
\ETotBD(R)=E_{pin}(R)+E_{\K}(R)\, ,
\eeq
with $E_{pin}$ the pinning energy and $E_\K$ the bending energy. In cylindrical coordinates these are given respectively by :
\beq\label{Epin}
E_{pin}(R)\approx\gs \int_0^R \pi r dr\, ,
\eeq
\beq\label{EK}
\begin{split}
E_\K(R)&=\pi\K\int_0^R rdr \l[\l(\frac{1}{r}\partial_r+\partial_r^2\rg)h(r)\rg]^2\\
&-\pi\K\int_0^R rdr \l[\l(\frac{1}{r}\partial_r+\partial_r^2\rg)s(r)\rg]^2\, .
\end{split}
\eeq
In both these expressions the energy is measured from the totally pinned configuration. The total energy $\ETotBD(R)$ allows us to determine the stability of the solutions $R=0$ and \Eqref{radSol}. This is simply exemplified for the case of a parabolic well.

\subsection{Parabolic Well}
In the case of a parabolic profile $s(r)=S_0+\frac{S_2}{2} r^2$, we see that \eqref{radSol} implies that $h(r)\equiv s(r)$ and there is no solution for a partially detached membrane. If we allow for a quartic term $s(r)=S_0+\frac{S_2}{2} r^2+\frac{S_4}{24} r^4$ then we obtain that the detachment radius is given by:
\beq\label{parRsol}
R^2_*=\frac{3}{S_4}\sqrt{\frac{\gs}{\K}}
\eeq
where we have taken the minus sign in the second equation of \eqref{radSol} corresponding to the fact that the curvature of the detached membrane is smaller than the one of the substrate. Therefore $S_4$ needs to be a positive quantity for a solution to exist. From the total energy $\ETotBD(R)$ given by \Eqref{ETotBDDef}, it is easy to show that the solution given by \Eqref{parRsol} corresponds to a maximum of the energy profile and hence it is an unstable equilibrium solution while $R=0$ is a metastable minimum. $\ETotBD(R)\ra -\infty$ for $R\ra \infty$ and therefore $R_*$ signals the energy barrier for total depinning which is always the stable configuration. This is however a construction of the unbounded quartic profile we have chosen for the substrate. In the next sections we will study in detail a more physically sensible profile: a Gaussian depression or protrusion.

\subsection{Gaussian Depression}
\label{subsec:GaussDepBD}
A more realistic landscape for the substrate is the case of a Gaussian depression
\beq\label{GaussDep}
s(r)=G_s\l(1-e^{-\frac{r^2}{2\s^2}}\rg)\, .
\eeq
For this geometry, the curvature of the substrate varies from positive to negative along the radial coordinate and therefore we have to allow for both signs $\pm \sqrt{\frac{\gs}{\K}}$ in \Eqref{radSol}. However when the condition is applied to this particular shape we obtain:
\beq\label{expBC}
\frac{s'(R)}{R}-s''(R)=\frac{G_s}{\s^2}\frac{R^2}{\s^2}e^{-\frac{R^2}{2\s^2}}=\sqrt{\frac{\gs}{\K}}\, ,
\eeq
that is, only the positive sign leads to the existence of a solution since we have assumed $G_s>0$. The solution for the membrane's profile then is given by
\beq\label{expSol}
h(r)=\l\{\ba{ll}
 G_s-G_se^{-\frac{R^2}{2\s^2}}\l(1+\frac{R^2-r^2}{2\s^2}\rg)&\quad 0\le r\le R\\
s(r) \quad &\quad r> R\, ,
\ea\rg.
\eeq
with $R$ given by \eqref{expBC}.

\begin{figure}
 \centering
 \includegraphics[angle=270,width=8cm]{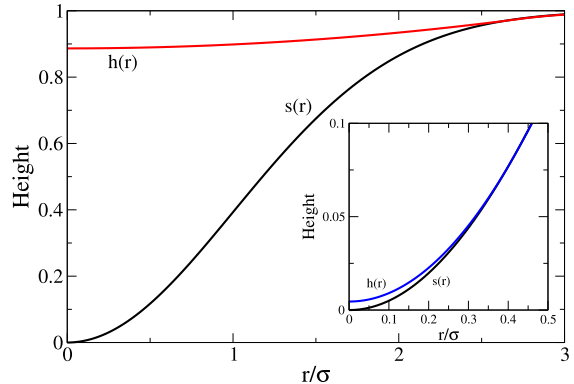}
\caption{Graphene depinning from a Gaussian depression in the BD regime. The plots show the profile of the graphene membrane $h(r)$ (Color) and of the substrate $s(r)$ (Black), in units of $\ml$, as a function of the distance from the center of the depression in units of the depression characteristic width $\s$. The figures have axial symmetry.  Main figure: stable solution with the graphene sheet almost completely detached ($R_2\approx 5.5$). Inset: Unstable solution with very little detachment of the graphene sheet ($R_1\approx 0.9$). Height profile as a function of the radial distance to the center of the depression, $r$. We have taken $G_s=1$ and $\s=2$. All quantities are in units of the characteristic length $\ml\approx 1$ \AA.}
 \label{fig:BigRSol}
\end{figure}

We apply now this solution to the particular case of graphene on top of a SiO$_2$ substrate. As we did in \Eqref{critlines}, in what follows we will treat all length quantities as dimensionless, given in units of the characteristic length $\ml$ defined in \Eqref{lenghtscale}. We can consider, as an example to illustrate the solutions given by \Eqref{expBC} and \Eqref{expSol}, a particular substrate depression of amplitude $G_s=1$ and width $\s=2$. We obtain two possible solutions for a partially detached configuration: $R_1\approx0.45$, and $R_2\approx2.75$.  In \figref{fig:BigRSol} we depict the graphene membrane profile solutions that correspond to this particular configuration.

\begin{figure}
 \centering
 \includegraphics[angle=270,width=8cm]{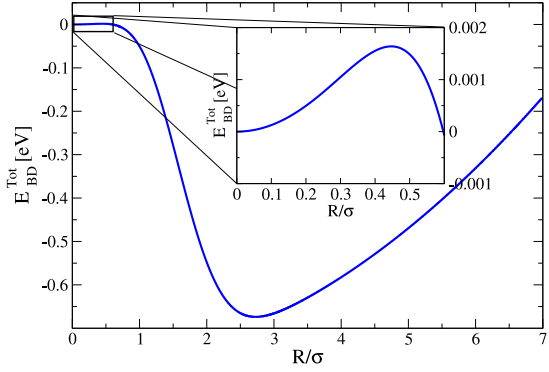}
\caption{Main figure: Total energy as a function of the detachment radius $R$ in units of the characteristic width $\s$, for a Gaussian depression in the bending rigidity dominated regime. Inset: Close up showing the metastable solution at $R_0=0$ and the unstable solution $R_1$. Results for $G_s=1$ and $\s=2$.}
\label{fig:ETotsmhdepG}
\end{figure}

To study the stability of the obtained solutions, we use the total energy which, by \Eqref{ETotBDDef} is given by ~\footnote{For the figures we have used the exact expression for the pinning energy, $E_{pin}=-\g\pi\int{ rdr\sqrt{1+\l(\frac{G_s}{\s^2}re^{-\frac{r^2}{2\s^2}}\rg)^2}}$ }:
\beq\label{EtotFlexDep}
\begin{split}
\ETotBD(R)&=\s^2\l\{\frac{\gs \pi  R^2}{2\s^2} \rg.\\&+\l.\frac{G_s^2}{\s^4}\K \pi\l[e^{-\frac{R^2}{\s^2}}\l(\half\frac{R^4}{\s^4}+\frac{R^2}{\s^2}+1\rg)-1\rg]\rg\} \,.
\end{split}
\eeq
The total energy $\ETotBD(R)$ corresponding to the substrate profile shown in \figref{fig:BigRSol}, is given in \figref{fig:ETotsmhdepG} as a function of the detachment radius $R$. We see that the completely pinned situation ($R_0=0$) is a metastable state with a very small energy barrier to overcome to reach the true minimum $R_2$. The solution $R_1$ corresponds to a maximum of the energy and hence it is an unstable configuration.

\begin{figure}
 \centering
 \includegraphics[angle=270,width=8cm]{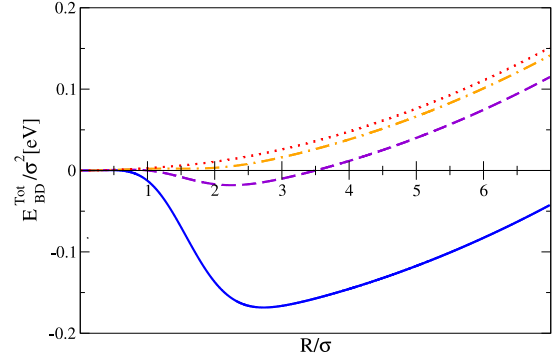}
\caption{(Color online) Energy landscape as a function of detachment radius in the bending rigidity dominated regime for a Gaussian depression with varying $\frac{G_s}{\s^2}$. Solid (blue): $\l.\frac{G_s}{\s^2}\rg|_1= 0.25$; dashed (purple): $\l.\frac{G_s}{\s^2}\rg|_2\approx 0.08$; dash-dot (orange): $\l.\frac{G_s}{\s^2}\rg|_3\approx 0.06$; dotted (red): $\l.\frac{G_s}{\s^2}\rg|_4\approx 0.03$. The energy presents a minimum for the detached configuration for cases $1$, $2$ and $3$ (being this last one metastable), while for case $4$ the energy is a minimum only for the completely pinned configuration, in agreement with the threshold value discussed in the main text. For cases $1$, $2$ and $3$ the completely pinned configuration is a local minimum with a low energy barrier, not visible due to the large scale of the plot. Note that $\s$ and $G_s$ are dimensionless: $\s\,,\,G_s \rightarrow\s/\ml\,,\,G_s/\ml$.}
 \label{fig:EpinToEKdep}
\end{figure}

By re-scaling the radius of detachment $R$ by the depression width $\s$, $\tilde{R}=R/\s$, the rescaled energy $\ETotBD\l(\tilde{R}\rg)/\s^2$ depends only on the ratio $G_s/\s^2$ and hence the results can be expressed in an universal manner. In \figref{fig:EpinToEKdep}, we show the re-scaled energy $\ETotBD\l(\tilde{R}\rg)/\s^2$ landscape for various values of $G_s/\s^2$. As expected, the global minimum corresponding to the partially detached configuration evolves into a metastable state as $G_s/\s^2$ is decreased, and disappears completely for small enough $G_s/\s^2$. As discussed in \secref{sec:Model}, the threshold value for a stable pinned configuration given by \Eqref{critlines} is $G_s/\s^2=\ml\sqrt{\frac{\gs}{\K}}\approx 0.05$. As it can be seen from the figure, this estimated threshold is in excellent agreement with the exact results.

\begin{figure}
 \centering
 \includegraphics[angle=270,width=8cm]{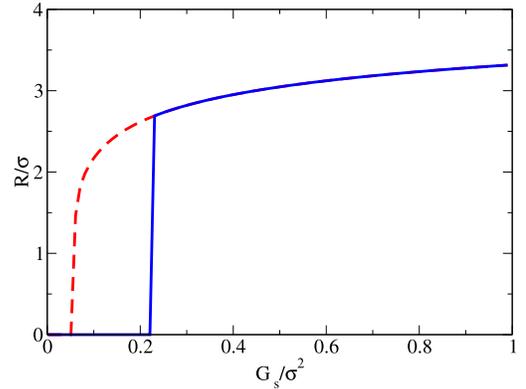}
 \caption{(Color online) Order parameter $\tilde{R}=R/\s$ as a function of the ratio $G_s/\s^2$ of a Gaussian depression in the BD regime. The dashed (red) line indicates the spinodal line, while the solid (blue) line corresponds to the true transition.}
 \label{fig:OrderParK}
\end{figure}

In \secref{sec:Model} we stated that the length of the detachment curve $\l|\bx^*\rg|$ is the natural order parameter that controls the pinned-to-depinned phase transition of the system. This can be easily seen now from \figref{fig:EpinToEKdep}. Given the cylindrical geometry of the problem, the length of the curve $\bx^*$ is given by $\l|\bx^*\rg|=2\pi R\ml$ and hence we can take $R$ as our order parameter. As discussed, from \figref{fig:EpinToEKdep} we see that the minimum at finite $R$ evolves into a metastable state that disappears for shallow enough depressions, while $R=0$ is the true minimum in this case, indicating that the transition is a first order one. This can be seen in an alternative way by following the evolution of the order parameter $R$. From \Eqref{expSol} again we note that $\tilde{R}$ is controlled solely by the ratio $G_s/\s^2$, in agreement with \Eqref{critlines} and with the universal form of the rescaled energy $\ETotBD\l(\tilde{R}\rg)/\s^2$. The behavior of $\tilde{R}$ as a function of $G_s/\s^2$ is shown in \figref{fig:OrderParK}, where we see that $\tilde{R}$ jumps from $\tilde{R}=0$ to a finite value at a critical value $\l.G_s/\s^2\rg|_c\approx 0.2$. The figure also shows the spinodal point, that is, the value $\l.G_s/\s^2\rg|_s\approx 0.05$ at which the first metastable solution appears, in agreement with the estimated threshold value. The difference between $\l.G_s/\s^2\rg|_c$ and $\l.G_s/\s^2\rg|_s$ shows that the system in the BD regime limit is strongly hysteretic.
The critical point $\l.G_s/\s^2\rg|_c$ and the spinodal point $\l.G_s/\s^2\rg|_s$ can be estimated from an expansion of the free energy \Eqref{EtotFlexDep} in the re-scaled order parameter $\tilde{R}$:
\beq\label{EBDexp}
\frac{\ETotBD}{\s^2}\approx \frac{\pi}{2}\l[\gs \tilde{R}^2+\K \l(\frac{G_s}{\s^2}\rg)^2\l(-\frac{\tilde{R}^6}{3}+\frac{\tilde{R}^8}{4}\rg)\rg]\,.
\eeq
This expansion can be identified with the Landau expansion \Eqref{LandauFreeEn} and it assumes that the re-scaled order parameter $\tilde{R}$ is small, and hence (assuming a weak first order transition) it is valid near the critical point $\l.G_s/\s^2\rg|_c$. By minimizing \Eqref{EBDexp} it is easy to see that the condition for the existence of metastable solutions with $\tilde{R}\neq0$ is given by $\l(G_s/\s^2\rg)^2\gtrsim\l(27/4\rg)\,\gs\approx0.12$, a value which is of the order of magnitude of the spinodal point obtained exactly in \figref{fig:OrderParK}. For $\l(G_s/\s^2\rg)^2\l(\gtrsim27/4\rg)\,\gs$, the extrema condition $d \ETotBD/dR=$ 0 in expression \Eqref{EBDexp} renders $\tilde{R}_0=0$ plus two real positive roots in agreement with the energy profiles presented in  \figref{fig:EpinToEKdep} for the exact solution \Eqref{EtotFlexDep}.

\subsection{Gaussian bump}
\label{subsec:GaussBumpBD}
For a Gaussian {\it protrusion}
\beq\label{bump}
s(r)=G_s e^{-\frac{r^2}{2\s^2}}
\eeq
the general solution \Eqref{GenSolh} holds for $r>R$. Since the origin is avoided, $H_1$ and $H_3$ can be different from zero. This solution however diverges for $r\ra \infty$ unless $h(r)=const$, which in turn cannot satisfy $h^\prime(R)=s^\prime(R)$. Hence a kind of solution for which the graphene membrane follows the substrate for $0<R<r$ and then detaches ``forever'' is not possible. The most general solution is to assume that there is a radius of detachment $R$ and a radius of re-attachment $L$, with $R<L$. The gain in pinning and bending energies with respect to the totally attached configuration in this case are given by:
\beq\label{EpinBump}
E_{pin}(R)\approx\gs \int_R^L \pi r dr\, ,
\eeq
\beq\label{EKBump}
\begin{split}
E_\K(R)&=\pi\K\int_R^L rdr \l[\l(\frac{1}{r}\partial_r+\partial_r^2\rg)h(r)\rg]^2\\
&-\pi\K\int_R^L rdr \l[\l(\frac{1}{r}\partial_r+\partial_r^2\rg)s(r)\rg]^2\, .
\end{split}
\eeq
\begin{figure}
 \centering
 \includegraphics[width=8cm]{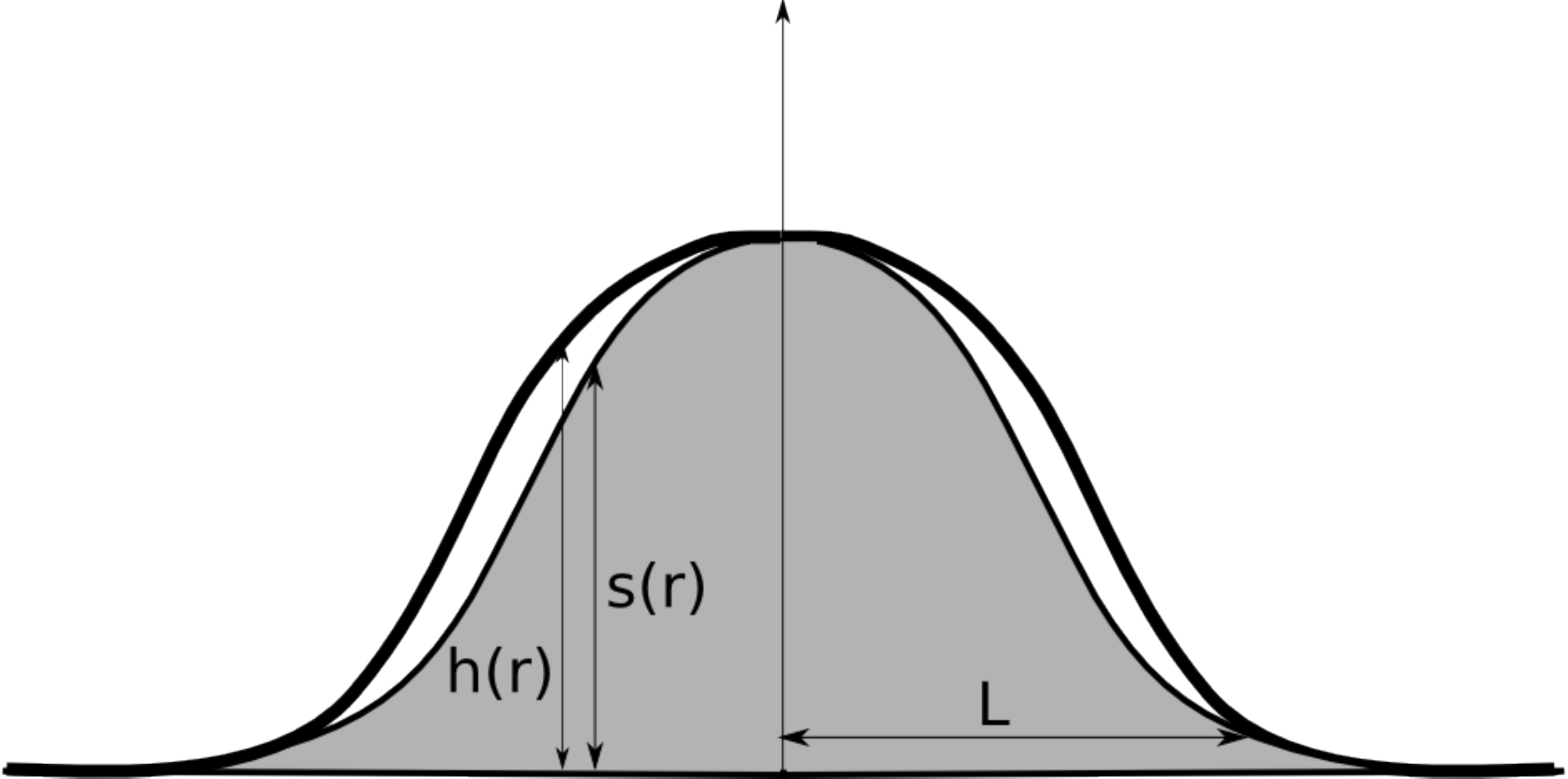}
 \caption{Membrane on top of a substrate with a protrusion. The figure is axially symmetric with respect to the vertical axis through the center of the substrate's bump. As discussed in the text, the membrane is shown as depinning from the top of the bump, and re-attaching at a radius L.}
 \label{fig:GrapheneOnGaussBump}
\end{figure}

\begin{figure}
 \centering
 \includegraphics[angle=270,width=8cm]{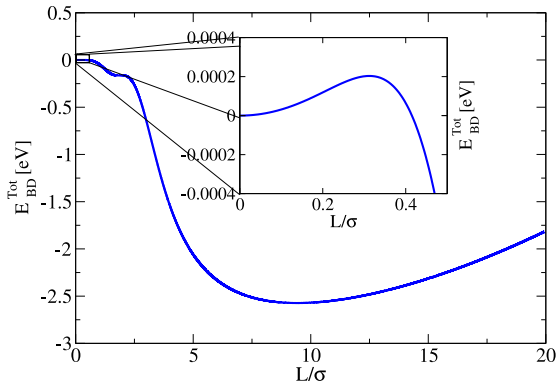}
\caption{Main figure: Total energy as a function of the re-attachment radius $L$ for a pronounced Gaussian bump in the bending rigidity dominated regime. Inset: Close-up near the metastable configuration corresponding to a completely pinned membrane. We have taken $G_s=1$ and $\s=2$.}
 \label{fig:FlexBump}
\end{figure}

Intuitively it is easy to see that if the bump is very pronounced, $R$ has to be approximately zero, otherwise the bending energy cost is too high. A sketch of the system is shown in \figref{fig:GrapheneOnGaussBump}. Assuming this type of configuration, the general solution \Eqref{GenSolh} holds now for $0\leq r\leq L$, with $H_1=0$ for it to be regular at the origin. The constant $H_3$ in this case is allowed to be finite since the contact force is acting at $r=0$ ~\cite{LandauElasticityBook}. Imposing continuity of the solution and its first derivative at $r=L$, and $h(0)=s(0)$ (note that $h(r)\ra H_0$ for $r\ra 0$), for the region $0\le r\le L$ we obtain for the membrane profile:
\begin{widetext}
\beq\label{expSolhbump}
h(r)\!= G_s\!-\!\frac{ G_s}{L^2}\!e^{-\frac{L^2}{2 \s^2}}\! \left[\!-\!1\!+\!e^{\frac{L^2}{2 \s^2}}\!-\!\frac{L^2}{2\s^2} \log{(L^2)}\!-\! \log{(L^2)}\!+\! e^{\frac{L^2}{2 \s^2}} \log{(L^2)}\right] r^2+\frac{G_s}{L^2} \!e^{-\frac{L^2}{2 \s^2}} \!\left[-\frac{L^2}{2\s^2}\!-\!1\!+\! e^{\frac{L^2}{2 \s^2}} \right]r^2\log{(r^2)}\, ,
\eeq
while $h(r)=s(r)$ for $r> L$. The optimal value of $L$ can be obtained numerically, as previously, by imposing the discontinuity of the Laplacian of the solution due to the contact force. However since we are interested in the qualitative aspect of the solution, it is simpler to analyze directly the energy profile as a function of the re-attachment radius $L$. As it was the case for the Gaussian depression in \subsecref{subsec:GaussDepBD}, the energy \Eqref{EnTotBDbump} and the re-attachment radius $L$ can be rescaled by the width of the bump to show the universal behavior.
\beq\label{EnTotBDbump}
\frac{\ETotBD(\tilde{L})}{\s^2}=\gs\pi\frac{\tilde{L}^2}{2}+\K \pi\frac{G_s^2}{2\s^4}e^{-\tilde{L}^2}\l[\tilde{L}^4+6\tilde{L}^2+2\l(9-8e^{-\frac{\tilde{L}^2}{2}}\rg)-16\l(1-e^{-\frac{\tilde{L}^2}{2}}\rg)^2\frac{1}{\tilde{L}^2}\rg]-\K \pi\frac{G_s^2}{2\s^4}\,,
\eeq
\end{widetext}
with $\tilde{L}=L/\s$. The energy profile \Eqref{EnTotBDbump} is shown in \figref{fig:FlexBump} for a bump with $G_s=1$ and $\s=2$, as a function of the re-attachment radius $\tilde{L}$. From the figure, it can be seen that the case of total adhesion of the graphene membrane to the substrate, in the case of a pronounced bump, is a metastable state with a very low energy barrier to fall into a configuration for which the membrane attaches to the substrate after a finite radius $L$.

\begin{figure}
 \centering
 \includegraphics[width=8cm]{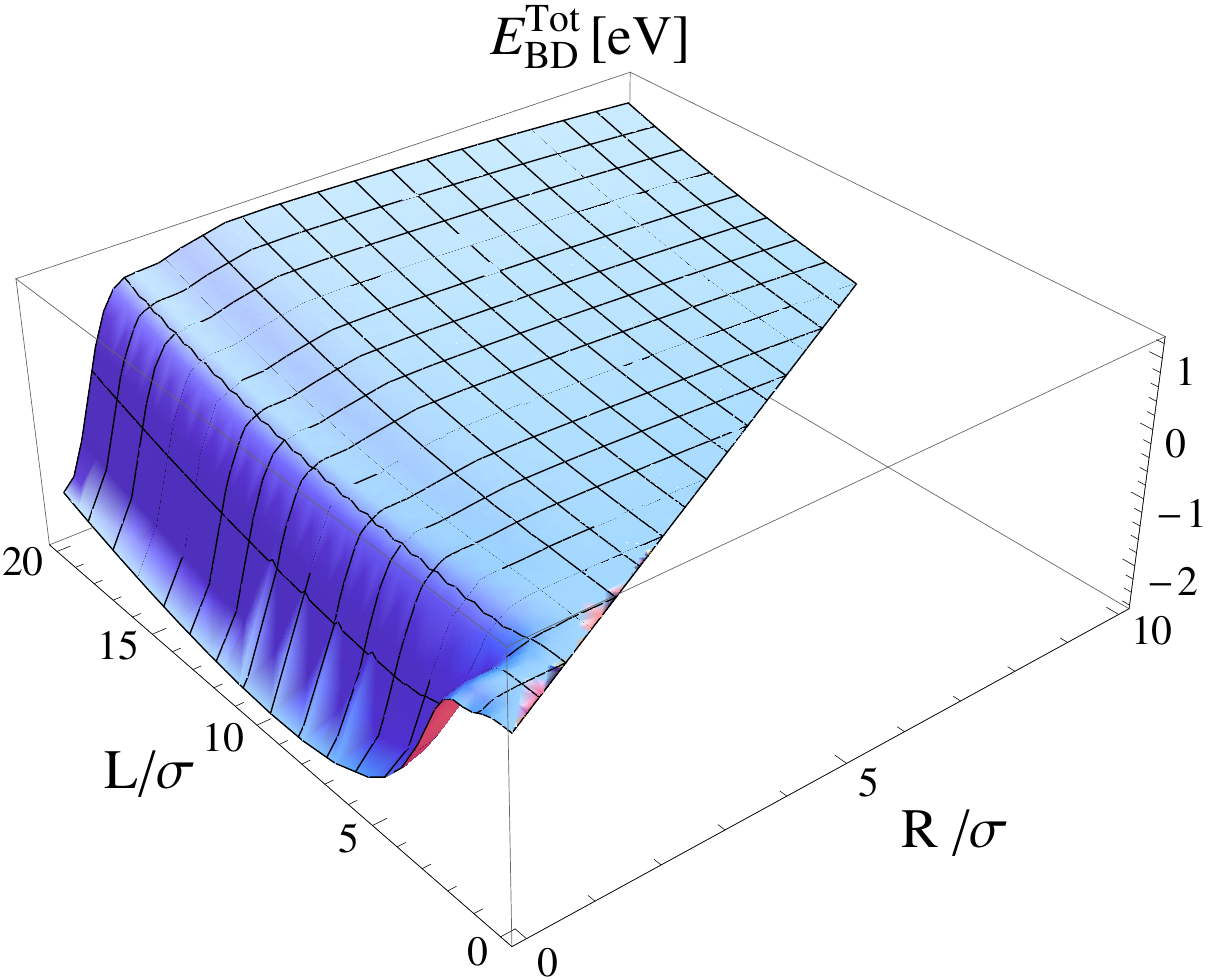}
\caption{Energy profile as a function of detachment radius $\tilde{R}=R/\s$ and re-attachment radius $\tilde{L}=L/\s$ with $G_s/\s^2=1$ for a Gaussian bump. The global minimum corresponds to $\tilde{R}=0$ and finite $\tilde{L}$.}
 \label{fig:FlexBump3D}
\end{figure}

\begin{figure}
 \centering
 \includegraphics[angle=270,width=8cm]{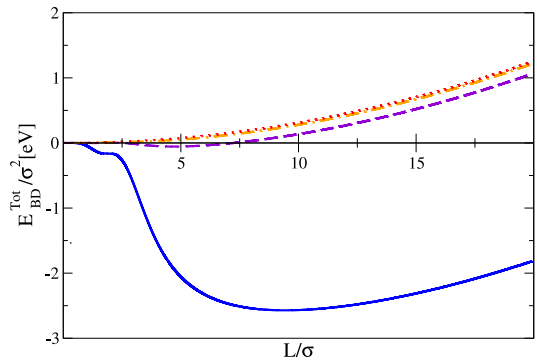}
 \caption{(Color online) Energy landscape as a function of re-attachment radius $\tilde{L}=L/\s$ in the BD regime for a Gaussian protuberance of varying ratio $\frac{G_s}{\s^2}$. Solid (blue): $\l.\frac{G_s}{\s^2}\rg|_1= 1$; dashed (purple): $\l.\frac{G_s}{\s^2}\rg|_2= 0.25$; dash-dot (orange): $\l.\frac{G_s}{\s^2}\approx 0.1\rg|_3$; dotted (red): $\l.\frac{G_s}{\s^2}\approx 0.04\rg|_4$. The energy presents a minimum for the partially detached configuration for case $1$ which disappears completely for case $4$, going through a metastable state for case $3$. Note that $\s$ is dimensionless: $\s \rightarrow\s/\ml$.}
 \label{fig:EpinToEKbump}
\end{figure}

As we mentioned previously, the solution \Eqref{expSolhbump} is valid in principle for pronounced Gaussian protrusions, for which $G_s/\s^2\gtrsim1$. However it can be shown that this is true for any Gaussian bump. This can be seen more rigorously by calculating the most general solution for which both the depinning and re-attachment radius are finite, and finding the minimum of the energy surface. The explicit solution for this most general case is rather cumbersome and it is given in \appref{sec:App1}, here we show a plot of the energy surface profile as a function of both detachment and re-attachment radius $\tilde{R}$ and $\tilde{L}$. As it can be seen from \figref{fig:FlexBump3D}, the complete solution indeed shows that the case $R=0$ and finite $L$ is a minimum for the case $G_s/\s^2=1$, and the same can be shown for other aspect ratio protrusions.
From \Eqref{EnTotBDbump} it is evident that $\ETotBD\l(L/\s\rg)/\s^2$ depends only on the ratio $G_s/\s^2$.
The re-scaled energy profile projection onto the $R=0$ plane, $\ETotBD(\tilde{L})/s^2$, for varying $G_s/\s^2$ is shown in \figref{fig:EpinToEKbump}, showing the crossover from the pinned to the partially detached configuration for increasingly pronounced bumps. As in the case of a Gaussian depression analyzed in the previous subsection, the case of a finite re-attachment radius $\tilde{L}$ is the energy minimum for $G_s/\s^2\gtrsim 0.05$, while in the opposite limit the minimum corresponds to $L=0$, that is, for smooth bumps the membrane minimizes its energy by conforming completely to the substrate.

Similar results to those obtained in \subsecref{subsec:GaussDepBD} for the behavior of the re-scaled order parameter $\tilde{R}$ as a function of $G_s/\s^2$ can be obtained here for $\tilde{L}$, showing a first order phase transition between the pinned and de-pinned phases.

\section{Detachment due to in-plane modes for radial symmetry}
\label{sec:InPlaneModes}
In the previous sections we studied the detachment of a graphene membrane from a patterned substrate due to the bending rigidity term in the free energy \Eqref{FreeEnEffTot}. This approximation is widely used, but, as we showed, it is valid for relatively small fluctuations of the substrate landscape for a 3D pattern. In this section we consider the less studied case of depinning due to in-plane modes, for which the free energy \Eqref{FreeEn} is approximated by \Eqref{IPFreeEn}. As before, we will restrict our study to cases that allow for an analytic solution, in particular a substrate with radial symmetry. In cylindrical coordinates this elastic energy is given by:
\beq\label{Eel}
\begin{split}
E_{el}&=\frac{\lam}{2}\int 2\pi r dr\l[\partial_r u_r+\frac{u_r}{r}+\half \l(\partial_r h\rg)^2\rg]^2\\&+\mu\int 2\pi r dr\l[\partial_r u_r+\half \l(\partial_r h\rg)^2\rg]^2 +\mu\int 2\pi r dr\l(\frac{u_r}{r}\rg)^2\, ,
\end{split}
\eeq
where $u_r$ is the radial component of the in-plane displacements, and $E_{pin}$ was defined in \eqref{Epin}. In the following subsections we will obtain results for a Gaussian depression and protrusion.

\subsection{Gaussian Depression}
\label{subsec:GaussDepSD}
In this section we consider again a Gaussian depression given by \Eqref{GaussDep}, over which there is a membrane partially attached. For radius greater than a radius $R$ the membrane is pinned to the substrate and follows its profile while for $0<r<R$ the membrane is completely detached. When analyzing the effect of the in-plane modes we encounter an added complication which is that we do not know the height profile of the membrane for the detached region, since this would imply to solve the problem completely by treating the full coupled non-linear differential equations in both $h$ and $\bu$ fields resulting from minimizing \Eqref{FreeEn}. Here we consider as a first approximation that the graphene membrane remains flat within the detached region as shown in \figref{fig:GrapheneOnGaussHole}. Hence the differential equation to solve is given by:
\begin{multline}\label{difEqrR}
-(\lam+2\mu)\l(\partial_r^2 u_r+\frac{\partial_ru_r}{r}-\frac{u_r}{r^2}\rg)=\\
\l\{\ba{ll}
0 \quad &\quad 0\le r\le R\\
(\lam+2\mu)\partial_r s \l(\partial_r^2 s\rg)+ \frac{\mu}{r}\l(\partial_r s\rg)^2 \quad &\quad r> R
 \ea\rg.
\end{multline}
Our ansatz corresponds to a membrane profile given by:
\beq\label{GaussProf}
h(r)=\l\{\ba{ll}
h_0 &\quad 0\le r\le R\\
s(r) \quad &\quad r> R
\ea\rg.
\eeq
with $s(r)$ given by \Eqref{GaussDep}. The general solution of \Eqref{difEqrR} is given by (see App. \ref{sec:App2}):
\begin{widetext}
\beq\label{ugralsol}
u_r(r)=\l\{\ba{ll}
r\frac{G_s^2}{4\s^2}\frac{\mu}{(\lam+2\mu)} e^{-\frac{R^2}{\s^2}} &\quad 0\le r\le R\\
\frac{G_s^2}{4}e^{-\frac{r^2}{2\s^2}}\l[\frac{r}{\s^2}+\frac{(\lam+\mu)}{(\lam+2\mu)}\frac{1}{r}\rg]
-\frac{1}{r}\frac{G_s^2}{4\s^2}\frac{(\lam+\mu)}{(\lam+2\mu)}\l(R^2+\s^2\rg) e^{-\frac{R^2}{\s^2}}  \quad &\quad r > R\, .
\ea\rg.
\eeq
The radius $R$ of detachment can be found by finding the extrema of the total energy $\ETotSD(R)=E_{pin}(R)+E_{el}(R)$ where $E_{pin}$ is given by \Eqref{Epin} and we measure the elastic energy \Eqref{Eel} from the totally attached configuration. The total energy as a function of detachment radius can be calculated to be:
\beq\label{EelDep}
\ETotSD(R)=\s^2\frac{\pi}{2}\l\{\gs\frac{2 R^2}{\s^2}+\frac{G_s^4}{\s^4}\frac{\mu (\lam+\mu) \pi }{8 (\lam+2 \mu) }\l[ e^{-\frac{2 R^2}{\s^2}}   \left(2\frac{R^2}{\s^2}+1\right)-1\rg]\rg\}\,.
\eeq
\end{widetext}
Minimizing $\ETotSD(R)$ renders a solution with $R_0=0$ which is a local minimum and corresponds to the membrane completely pinned, and the following transcendental equation for the equilibrium detachment radius:
\beq
\gs=\frac{G_s^4}{2\s^4}\frac{R^2}{\s^2}\mu^2\frac{(\lam+\mu)}{(\lam+2\mu)^2}e^{-\frac{2 R^2}{\s^2}}\,.
\eeq
Again, as in the BD dominated regime case of \secref{sec:FlexModes}, we see that by re-scaling both the detachment radius and the total energy by an overall factor given by the depression width $\s$, $\tilde{R}=R/\s$ and $\ETotSD(\tilde{R})/\s^2$, the re-scaled energy shows universality. In this case, and in agreement with \Eqref{critlines}, the system is controlled by the ratio $G_s/\s$, opposed to the dependence on $G_s/\s^2$ found for the BD regime.

\begin{figure}
 \centering
 \includegraphics[angle=270,width=8cm]{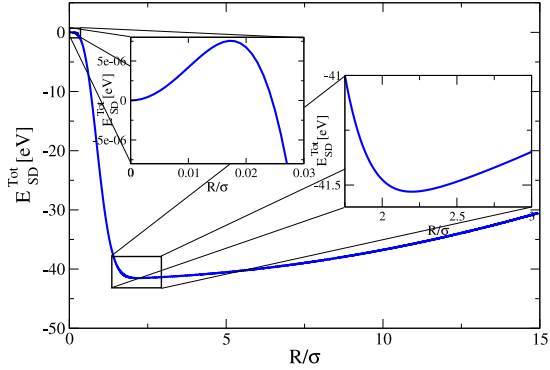}
\caption{Energy profile for a Gaussian depression as a function of detachment radius $\tilde{R}=R/\s$ for $G_s=5$ and $\s=4$ as discussed in the main text. The insets show the local minimum that corresponds to the totally pinned configuration, and the global minimum corresponding to the partially detached membrane.}
 \label{fig:EnProfileGDep}
\end{figure}

We can apply our results to a graphene membrane on top of a SiO$_2$ substrate as we did in \secref{sec:FlexModes}.
Taking the accepted values for room temperature for the Lam\'{e} coefficients of graphene, $\mu \approx 10$ eV \AA$^{-2}$, $\lam \approx 2$ eV \AA$^{-2}$ ~\cite{Fasolino09}, and $G_s=5$ (in accordance with the validity of our approximation) and $\s=4$ (note that we are still working in in units of the scaling length $\ml$) we get two possible depinning radius, $R_1\approx 0.1$ corresponding to an unstable minimally detached configuration,  and $R_2 \approx 8.4$ which is the stable, global minimum solution. The total energy $\ETotSD$ is plotted as a function of detachment radius $\tilde{R}=R/\s$ in \figref{fig:EnProfileGDep}, showing the different equilibrium solutions.The energy barrier to be overcome for detachment from the metastable equilibrium configuration at $R_0=0$ is very small as can be seen from the inset in \figref{fig:EnProfileGDep}. Note also that the minimum at $R_2$ is very shallow in comparison to the energy scale, as shown in the inset of \figref{fig:EnProfileGDep} and hence any fluctuation could lead to the graphene membrane to be detached at a radius $R>R_2$, and therefore closer to the flat configuration. This effect is less pronounced as the width of the depression is increased.

\begin{figure}
 \centering
 \includegraphics[angle=270,width=8cm]{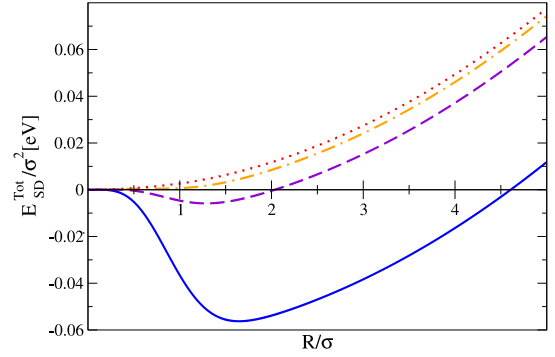}
 \caption{(Color online) Energy landscape in the SD regime as a function of detachment radius $\tilde{R}=R/\s$, for a Gaussian depression of varying ratio $\frac{G_s}{\s}$. Solid (blue): $\l.\frac{G_s}{\s}\rg|_1= 0.5$; dashed (purple): $\l.\frac{G_s}{\s}\rg|_2\approx 0.33$); dash-dot (orange): $\l.\frac{G_s}{\s}\rg|_3=0.25$; dotted (red): $\l.\frac{G_s}{\s}\rg|_4\approx 0.17$. The energy presents a minimum for the detached configuration for cases $1$, $2$ and $3$ (being this last one metastable), while for case $4$ the energy is a minimum only for the completely pinned configuration, in agreement with the threshold value discussed in the main text. For cases $1$ and $2$ the completely pinned configuration is a local minimum with a low energy barrier, not visible due to the large scale of the plot. The region of metastability for the depinned configuration is very small, as seen in this plot and also in \figref{fig:OrderParS}. Note that $\s$ is dimensionless: $\s \rightarrow\s/\ml$.}
 \label{fig:InPlaneMultiEnDep}
\end{figure}

In \secref{sec:Model}, \Eqref{critlines}, for the SD regime we predicted a critical value for the ratio $G_s/\s$ above which the stable configuration is that of the membrane partially detached from the substrate. Taken the values for graphene discussed above, this ratio is given by $\l(4\frac{\gs}{\Etd}\rg)^{1/4}\approx0.1$. This estimate is in good agreement with the exact results for the re-scaled energy profile $\ETotSD(\tilde{R})/\s^2$ depicted in \figref{fig:InPlaneMultiEnDep}, as a function of the re-scaled detachment radius $\tilde{R}$ and varying $G_s/\s$.

\begin{figure}
 \centering
 \includegraphics[angle=270,width=8cm]{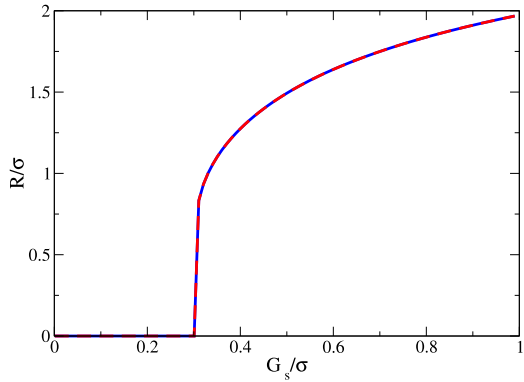}
 \caption{(Color online) Order parameter $R$ as a function of height $G_s$ of the Gaussian depression in units of $\s$ in the SD regime. The dashed (red) line indicates the spinodal line, while the solid (blue) line corresponds to the true transition. In the SD regime these two lines are indistinguishable. }
 \label{fig:OrderParS}
\end{figure}

As discussed in \secref{sec:FlexModes}, the radius of detachment $R$ can be taken as the order parameter of the problem. We show the behavior of $\tilde{R}=R/\s$ as a function of $G_s/\s$ in \figref{fig:OrderParS}. Again, a pronounced jump in $\tilde{R}$ to a finite value with increasing $G_s/\s$ is observed, at a critical value $\l.G_s/\s\rg|_c$ in agreement with the predicted threshold for the transition. The spinodal line, also shown in \figref{fig:OrderParS}, is basically indistinguishable from the true transition and hence in the SD regime there is almost no hysteresis.
The expansion of the free energy \Eqref{EelDep} in powers of the order parameter $\tilde{R}$:
\beq\label{ESDexp}
\frac{\ETotSD\l(\tilde{R}\rg)}{\s^2}\approx \frac{\pi}{2}\l\{\gs \tilde{R}^2 + \frac{G_s^4}{\s^4}\frac{\mu\l(\lam+\mu\rg)}{\l(\lam+2\mu\rg)}\l[-\frac{\tilde{R}^4}{4}+\frac{\tilde{R}^6}{3}\rg]\rg\}
\eeq
gives a good qualitative description of the first order transition obtained exactly in \figref{fig:OrderParS}. Moreover, in this case the spinodal point given by the expansion \Eqref{ESDexp} is given by $\l(G_s/\s\rg)^4=\l(44/15\rg) \gs$, giving $\l.G_s/\s\rg|_s\approx 0.28$, in excellent numerical agreement with the exact value.

\subsection{Gaussian bump}
\label{subsec:GaussBumpSD}
\begin{figure}
 \centering
 \includegraphics[angle=270,width=8cm]{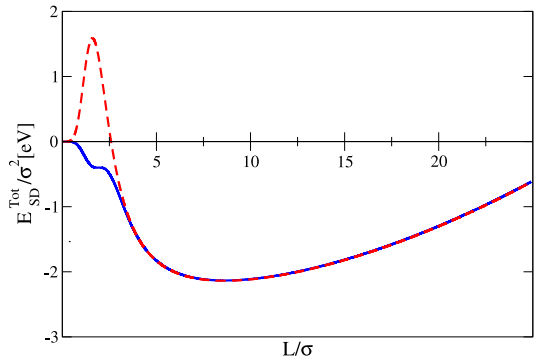}
 \caption{(Color online) The two possible solutions for the energy profile as a function of re-attachment radius $\tilde{L}=L/\s$ in the SD regime, as calculated in \appref{sec:App2}. We have taken $G_s=5$ and $\s=4$.}
 \label{fig:EnBumpSD}
\end{figure}

Following similar manipulations to the previous section, we can calculate the solution for the radial component for the in-plane displacements in the graphene membrane due to a Gaussian protrusion parametrized by \Eqref{bump}. Given our findings for the BD regime in \subsecref{subsec:GaussBumpBD}, we consider a configuration of membrane on top of the substrate that is pinned at the very top (detachment radius $R=0$) and re-attaches at a radius $L$ as shown in \figref{fig:GrapheneOnGaussBump}. This takes into account the energetic cost of bending. As in \subsecref{subsec:GaussDepBD}, we have to make a sensible approximation for the unknown profile of the membrane on the detached section. We hence approximate the detached profile of the membrane by the general solution valid for small protrusions (BD regime), \Eqref{GenSolh} ~\cite{TimoshenkoBook}.
The details for the solution using this ansatz are given in \appref{sec:App2}. Imposing the boundary conditions results in two possible solutions, leading to the two energy profiles shown in \figref{fig:EnBumpSD} for $G_s=5$ and $\s=4$. Although these solutions differ for the metastable or unstable regions (a construction of the approximation involved), they coincide for the minimum and hence the stable re-attachment radius $L$ is uniquely defined.

\section{Discussion}
\label{sec:Disc}

In this work we have analyzed the possibility of depinning of a membrane on top of a patterned substrate. We have studied simple configurations of the substrate that allow for analytical solutions, in the two relevant limits for the problem: the bending rigidity dominated regime, valid for small corrugations of the substrate, and the elastic, strain dominated regime which holds for larger corrugations. We have shown that in both limits, the energy cost of either bending or stretching can cause the membrane to prefer to detach from the substrate. The particular results confirm the more general picture we sketched in \secref{sec:Model}, in which we obtained a qualitative phase diagram for the system presented in \figref{fig:phasediag}. This phase diagram presents two phases, a completely pinned phase in which the membrane follows the profile of the substrate, and a detached phase in which the membrane prefers to depin partially at some optimal detachment curve $\bx^*$. We have shown that an order parameter can be constructed from this detachment curve which allows us to write the problem in terms of a Landau energy functional \Eqref{LandauFreeEn}. By scaling arguments we were also able to obtain the critical lines \eqref{critlines} in both BD and SD limiting regimes and found excellent agreement with analytical calculations. The critical line in \figref{fig:phasediag} represents a first order phase transition, as argued in \secref{sec:Model} and shown explicitly for specific geometries of the substrate. We showed that the energy of the system shows universality, and depends on the ratio of $S/L$ or $S/L^2$ for the SD and BD regimes, where $S$ is the height of the substrate's corrugation and $L$ its characteristic width. We obtained the critical and spinodal points exactally for the analytically solvable cases,  and showed that the Landau energy functional gives a very good estimate to locate these points.

The depinning process is dependent on the aspect ratio of the spatial perturbations of the substrate, the elastic parameters of the membrane, and its interaction with the substrate. The interaction between graphene and different
types of substrates is not well known. Order of magnitude estimates
for different mechanisms ~\cite{Sabio08} suggest that the interaction coupling $\gs \sim
10^{-2} - 2$ meV \AA$^{-2}$ (note that the interaction between two
graphene layers in graphite is 20-30 meV \AA$^{-2}$ ~\cite{Sabio08}). Within this work we have set $\gs=2$meV \AA$^{-2}$, its most conservative value, and hence the obtained values for depinned configurations are underestimations. For this value of $\gs$ we find that, in the elastic regime, the
depinning of graphene becomes relevant for height corrugations such
that $S / L \gtrsim 1/10$. In the bending regime, the condition depends on the total area of the corrugation: $S / L^2 \gtrsim 1/20$ $ \text{\r{A}}^{-1}$. These obtained values for possible depinning are comparable to measured corrugations
in free standing graphene ~\cite{Meyer07}, and  in graphene on
SiO$_2$ ~\cite{Fuhrer07,Stolyarova07}. Hence, regions where graphene is
detached from the substrate may be found in samples on SiO$_2$, in
agreement with the observations reported in\cite{Morgenstern09}. Although these conditions were obtained for the two limiting regimes, in real life both effects are present. In general, corrugations of all scales are ubiquitous due to the intrinsic roughness of the substrate, and both the bending rigidity and the in-plane strains of the graphene membrane will contribute to its depinning. Random configurations of the substrate within a mean field model as presented in \secref{sec:Model} could be treated by adding noise to the system, considering the parameters in the Landau free energy \Eqref{LandauFreeEn} as random, in the spirit of random mass theories.

The presence of corrugations in graphene, either intrinsic or substrate-induced, can lead to diverse experimental consequences. Corrugations in graphene are associated with gauge fields that couple to the Dirac electrons \cite{AntonioRMP}. These gauge fields generate an effective magnetic field that can affect the transport properties of graphene. We have shown however that, for graphene, the scale of the corrugations for which the bending rigidity is relevant is rather small. As we saw in \secref{sec:FlexModes}, the BD to SD regime crossover length for graphene is $\ml\approx1\,$ \AA and hence depinning due to the strain energy cost is to be expected. The effective magnetic field is related to the strain through the relation
\beq\label{effBfield}
B \sim \f_0 \frac{\b}{a}\frac{\a}{L}\, ,
\eeq
where $\f_0 \approx 10^{-15}$ Wb is the quantum of magnetic flux, $a=1.42\,$ \AA $\,$ is the lattice parameter and $\b\approx 2$ gives the change in the hopping parameter between nearest sites for a Dirac electron due to the deformation of the lattice. The corrugation of the membrane determines $L$, the characteristic width of the corrugation, and the strain $\a$, which for simplicity we assume to be constant. We can roughly estimate the maximum magnetic field that the graphene membrane can experience do to the corrugation of the substrate. The strain of a corrugation of height $s$ and width $L$ scales as $s^2/L^2$ and hence, from  \Eqref{thresholdEl}, the maximum strain that graphene can support is $\a_m\approx 2\%$. As we showed, beyond this point the membrane relaxes by depinning partially from the substrate and lowering the strain. Taking a physically relevant corrugation width of $L\approx 100$ nm, $\a_m$ corresponds to a maximum effective magnetic field of $B_m\approx2$ T. This order of magnitude indicates that, indeed, the magnetic field due to the induced corrugations can have a sizeable effect in the transport properties of graphene ~\cite{Morozov06,Prada09}. Our results indicate that rougher substrates could in fact lead to flatter configurations of the graphene membrane after annealing, due to the impossibility of graphene to conform to pronounced depressions or bumps. This could result, in a counterintuitive fashion, in greater mobilities for graphene on top of very rough substrates, due to a decrease in impurity and phonon scattering \cite{Hwang07,Fuhrer08,Bolotin08,Morozov08}. On the other hand, for these kind of substrates the graphene membrane would be almost suspended and therefore prone to the excitation of flexural modes which contribute to the resistivity~\cite{Guinea10,vonOppen10}. A more unexplored path is the possibility of controlling the pinning or not of graphene to the substrate by tuning the different metastable states, which could be realized by modulating the gate electric field or by applying external pressure. As we showed, the system can present hysteresis and irreversibility. Our results can be also helpful to the understanding of the ubiquitous formation of graphene bubbles on many types of substrates\cite{Setal09,Letal10,K10,A10}.

To conclude, we list the limitations of our model. Our work is based on a continuum approach and hence it breaks down for lengths of the order of the lattice spacing. However amplitudes of the order of the lattice spacing can still be well described by the continuum model as shown in Ref. \onlinecite{Los09}. As we pointed out, the phase diagram presented in \figref{fig:phasediag} is valid for 3D profiles of the substrate. The crossover to 1D geometries would in principle imply the disappearance of the linear critical line valid for the SD regime, since the BD results are exact for 1D. The same holds for exact calculations of \secref{sec:FlexModes} and \secref{sec:InPlaneModes}, which have been done for a isotropic perturbations. The effect of the lack of radial symmetry remains to be explored. Lastly, it is possible that the interaction between the graphene layer and the substrate is not uniform, due to the presence of charges and
other defects within the substrate. The modeling of this kind of potential goes beyond the scope of this paper.

\acknowledgements
S. V. K. thanks Alex Kitt for fruitful conversations and William Cullen for relevant suggestions. A. H. C. N. acknowledges DOE grant DE-FG02-08ER46512 and ONR grant MURI N00014-09-1-1063. F. G. acknowledges financial support from MICINN (Spain), Grants FIS2008-00124 and CONSOLIDER CSD2007-00010.

\appendix
\begin{widetext}
\section{Full solution for a Gaussian bump: flexural modes}
\label{sec:App1}
In the main text we have analyzed the stability of a membrane on top of a Gaussian bump, which is pinned at the very top of the bump an re-attaches to the substrate at a radius $L$. This was an assumption based on energetic arguments. The most general solution can have the membrane conforming to the substrate from the top up to a finite depinning radius $R$, and then re-attaching at a radius $L$. In this case the general solution \Eqref{GenSolh} is valid for $R<r<L$ and all the coefficients $H_i$, $i=0,...3$ can be finite. These can be found by imposing the continuity of the solution and its first derivative both at $R$ and $L$. Defining 
$$\mathcal{H}=\frac{G_s e^{-\frac{L^2+R^2}{2 \s^2}}}{\s^2 \l[\left(L^2-R^2\right)^2-4 L^2 R^2 \l(\log{L}-\log{R}\rg)^2\rg]}$$ we have: 
\beq
\begin{split}
H_0&=\mathcal{H}\l[\l(L^2-R^2\rg) \l(e^{\frac{L^2}{2 \s^2}} L^2-e^{\frac{R^2}{2 \s^2}} R^2\rg) \s^2-2 e^{\frac{L^2}{2 \s^2}} L^2 R^2 \l(R^2+2 \s^2\rg) \log{L}^2\rg]\\
&+\mathcal{H}L^2 R^2\log{R} \l[2 e^{\frac{R^2}{2 \s^2}} \s^2+e^{\frac{L^2}{2 \s^2}} \l(L^2-R^2-2 \s^2\rg)-2 e^{\frac{R^2}{2 \s^2}} L^2 R^2 \l(L^2+2 \s^2\rg) \log{R}\rg]\\
&+\mathcal{H}L^2 R^2 \log{L}\l[2 e^{\frac{L^2}{2 \s^2}} \s^2+e^{\frac{R^2}{2 \s^2}} \l(-L^2+R^2-2 \s^2\rg)+2 \l(e^{\frac{R^2}{2 \s^2}} \l(L^2+2 \s^2\rg)+e^{\frac{L^2}{2 \s^2}} \l(R^2+2 \s^2\rg)\rg) \log{R}\rg]\\
H_1&= -\mathcal{H}L^2 R^2 \l[\left(e^{\frac{L^2}{2 \s^2}}-e^{\frac{R^2}{2 \s^2}}\right) \left(L^2-R^2\right)+\left(2 e^{\frac{R^2}{2 \s^2}} \left(L^2+2 \s^2\right)-2 e^{\frac{L^2}{2 \s^2}} \left(R^2+2 \s^2\right)\right) \log{L}\rg]\\
&+2\mathcal{H}L^2 R^2\l[e^{\frac{R^2}{2 \s^2}} \left(L^2+2 \s^2\right)-e^{\frac{L^2}{2 \s^2}} \left(R^2+2 \s^2\right)\rg]\log{R}\\
H_2&=-2\mathcal{H} \l[\l(e^{\frac{L^2}{2 \s^2}}-e^{\frac{R^2}{2 \s^2}}\rg) \l(L^2-R^2\rg) \s^2-2 e^{\frac{L^2}{2 \s^2}} L^2 R^2 \log{L}^2\rg]\\
&-2\mathcal{H}R^2\log{R}   \l[2 e^{\frac{R^2}{2 \s^2}} \s^2+e^{\frac{L^2}{2 \s^2}} \left(L^2-R^2-2 \s^2\right)-2 e^{\frac{R^2}{2 \s^2}} L^2\log{R}\rg]\\
&-2\mathcal{H}L^2\log{L} \l[2 e^{\frac{L^2}{2 \s^2}} \s^2+e^{\frac{R^2}{2 \s^2}} \l(-L^2+R^2-2 \s^2\rg)+2 \l(e^{\frac{L^2}{2 \s^2}}+e^{\frac{R^2}{2 \s^2}}\rg) R^2 \log{R}\rg]\\
H_3&=-\mathcal{H}\l(L^2-R^2\rg)\l[ e^{\frac{R^2}{2 \s^2}} \l(L^2+2 \s^2\rg)-e^{\frac{L^2}{2 \s^2}} \l(R^2+2 \s^2\rg)\rg]-2\mathcal{H}L^2R^2 \l[\l(e^{\frac{L^2}{2 \s^2}}-e^{\frac{R^2}{2 \s^2}}\rg)\log{L}- \l(e^{\frac{L^2}{2 \s^2}}-e^{\frac{R^2}{2 \s^2}}\rg) \log{R}\rg]
\end{split}
\eeq
The total bending energy can be calculated as in the main text taking into account that we now have three different regions of integration. The result, measured from the totally pinned configuration, is given by:
\beq
\begin{split}
E_\K&=\kappa\pi\mathcal{H}\frac{ (L^2 - R^2)}{2\s^2} \l[-16 e^{\frac{L^2 + R^2}{2 \s^2}}
          (L^2 R^2 + (L^2 + R^2) \s^2 + 2 \s^4)\rg] \\
&+ \kappa\pi\mathcal{H}\frac{ (L^2 - R^2)}{2\s^4}
       e^{\frac{R^2}{\s^2}} \l[L^6 - 2 \s^4 (R^2 - 8 \s^2) - L^4 (R^2 - 6 S^2) +
          2 L^2 \s^2 (R^2 + 9 S^2)\rg]\\
& + \kappa\pi\mathcal{H}\frac{ (L^2 - R^2)}{2\s^4}
       e^{\frac{L^2}{\s^2}} \l[R^6 + 2\s^2(3 R^4 + 9 R^2 \s^2 + 8 \s^4) -
          L^2 (R^4 - 2 R^2 \s^2 + 2 \s^4)\rg] \\
&+8 \kappa\pi\mathcal{H}\frac{L^2 R^2(\log{L} - \log{R})}{\s^2}  e^{\frac{L^2 + R^2}{2 \s^2}} (L^2 + R^2 + 4 \s^2)\\
& - 2\kappa\pi\mathcal{H}\frac{ L^2 R^2(\log{L} - \log{R})}{\s^4} e^{\frac{R^2}{\s^2}} \l[4 \s^2 (L^2 + 2 \s^2) + (L^4 + 2 L^2 \s^2 +
             2 \s^4)\l(\log{L}-\log{R}\rg) \rg] \\
&+2\kappa\pi\mathcal{H}\frac{ L^2 R^2(\log{L} - \log{R})}{\s^4} e^{\frac{L^2}{\s^2}} \l[-4 \s^2 (R^2 + 2 \s^2) + (R^4 + 2 R^2 \s^2 +
             2 \s^4) \l(\log{L}-\log{R}\rg)\rg]\,.
\end{split}
\eeq
This bending energy together with the contact energy cost gives the plot shown in \figref{fig:FlexBump3D}.

\section{Solution for a Gaussian depression and bump: In-plane modes}
\label{sec:App2}
\subsection{Gaussian depression}
We will call $u_r^<$ and $u_r^>$ the solutions of \eqref{difEqrR} for $0\le r\le R$ and $r>R$ respectively. For $0\le r\le R$ \eqref{difEqrR} is homogeneous and has a general solution of the kind:
$$
\frac{A_0}{r}+A_1 r\, .
$$
The constant $A_0\equiv0$ for the solution to be regular at the origin, and $A_1$ is determined by the boundary conditions.

For $r>R$ the homogeneous solution is given by
$$
\frac{C_0}{r}+C_1 r\, .
$$
but in this case we get $C_1\equiv0$ by imposing that the displacements are $0$ at infinity. $C_0$ is determined by boundary conditions. To obtain the general solution of \eqref{difEqrR} for $r>R$ we have to add a particular solution. This can be obtained by the ansatz solution:
\beq\label{ansatzu}
u_r=\sum_m a_m r^m e^{-\frac{r^2}{\s^2}}\, .
\eeq
Substituting \eqref{ansatzu} into the second line of \eqref{difEqrR} and using \eqref{GaussProf} we obtain:
\beq\label{polycond}
\sum_m a_m \l[\frac{4}{\s^4}r^{m+2}-\frac{4}{\s^2}(m+1)r^{m}+(m^2-1)r^{m-2}\rg]=\frac{G_s^2}{\s^6}r^3
-\frac{G_s^2}{\s^4}\frac{(\lam + 3\mu)}{(\lam + 2\mu)} r\, .
\eeq
From here we see that there is a possible solution with $m=\pm 1$. By substituting in \eqref{polycond} we obtain:
\beq
\label{coeffsa}
a_1= \frac{G_s^2}{4\s^2} \quad \quad \quad
a_{-1}= \frac{G_s^2}{4}\frac{(\lam +\mu)}{(\lam + 2\mu)}\, .
\eeq

The coefficients $A_1$ and $C_0$ are determined by imposing the continuity of the solution $u_r^>(R)=u_r^<(R)$ and of the in-plane stresses at $R$:
\beq\label{IPStress}
\s_{rr}=\lam\l[\partial_r u_r+\frac{u_r}{r}+\half \l(\partial_r h\rg)^2\rg]+2\mu\l[\partial_r u_r+\half \l(\partial_r h\rg)^2\rg]\, ,
\eeq
and hence the full solution is given by \eqref{ugralsol} in the main text.

\subsection{Gaussian bump}
As discussed in the main text, for a bump we approximate the membrane's profile in the detached region by the general solution given in the BD regime \Eqref{GenSolh}:
\beq\label{GenSolhApp}
h(r)=G_s+\frac{H_2}{2}r^2+H_3r^2\log{r}
\eeq
where we have set $H_0=G_s$ and $H_1=0$ as explained in \subsecref{subsec:GaussBumpBD}. If we denote the re-attachment radius as$L$, then the in-plane radial displacement $u_r$ is given by:
\beq\label{difEqurApp}
-(\lam+2\mu)\l(\partial_r^2 u_r+\frac{\partial_ru_r}{r}-\frac{u_r}{r^2}\rg)=
\l\{\ba{ll}
(\lam+2\mu)\partial_r h \l(\partial_r^2 h\rg)+ \frac{\mu}{r}\l(\partial_r h\rg)^2 \quad &\quad 0\le r\le L\\
(\lam+2\mu)\partial_r s \l(\partial_r^2 s\rg)+ \frac{\mu}{r}\l(\partial_r s\rg)^2 \quad &\quad r> L
 \ea\rg.
\eeq
together with the appropriate boundary conditions. The solutions for $0\le r\le L$ and $r > L$ are given respectively by:
\beq
\begin{split}\label{solSDbumpApp}
 u_r^<(r)=& C_1 r -  \frac{r^3\l[-2 H_1 H_2 (\lam + \mu) + 2 H_1^2 (\lam + 3 \mu) +
    H_2^2 (\lam + \mu) \rg]}{
  16 (\lam + 2 \mu)} \\&+
  \frac{r^3H_2 \log{(r)} \l[H_2 (\lam + \mu) - 2 \l(H_1+H_2 \log{(r)}\rg) (\lam + 3 \mu)\rg]}{
 4 (\lam + 2 \mu)}\\
u_r^>(r)=&\frac{C_0}{r}+\frac
  {G_s^2 e^{-\frac{r^2}{\s^2}}}{4 (\lam + 2 \mu) r} \l[\l(\lam + \mu\rg) + \l(\lam + 2 \mu\rg) \frac{r^2}{\s^2}\rg]\,.
\end{split}
\eeq
The parameters $H_1$, $H_2$, $C_0$ and $C_1$ are fixed by the boundary conditions, continuity of the solution \Eqref{solSDbumpApp} and its first derivative at $r=L$ and continuity of the in-plane stress  \Eqref{IPStress}, plus continuity of the flexural field $h(L)=s(L)$. Imposing these result in two possible sets of solutions:
\beq\label{CoeffSolSDbumpApp1}
\begin{split}
C_0=&-\frac{G_s^2 e^{-\frac{L^2}{\s^2}} (\lam +\mu ) \left(L^4+4 L^2 \s^2+8 \s^4-8 e^{\frac{L^2}{2 \s^2}} \s^4+4 e^{\frac{L^2}{\s^2}} \s^4\right)}{16 (\lam +2 \mu ) \s^4}\\
C_1=&\frac{G_s^2 e^{-\frac{L^2}{\s^2}} \mu  \left(L^4+5 L^2 \s^2-4 e^{\frac{L^2}{2 \s^2}} L^2 \s^2+8 \s^4-16 e^{\frac{L^2}{2 \s^2}} \s^4+8 e^{\frac{L^2}{\s^2}} \s^4\right)}{4 L^2 (\lam +2 \mu ) \s^4}\\
H_1=&\frac{2 G_s e^{-\frac{L^2}{2 \s^2}} \left(-\s^2+e^{\frac{L^2}{2 \s^2}} \s^2-L^2 \log{(L)} -2 \s^2 \log{(L)} +2 e^{\frac{L^2}{2 \s^2}} \s^2 \log{(L)} \right)}{L^2 \s^2}\\
H_2=&\frac{G_s e^{-\frac{L^2}{2 \s^2}} \left(-L^2-2 \s^2+2 e^{\frac{L^2}{2 \s^2}} \s^2\right)}{L^2 \s^2}
\end{split}
\eeq
\beq\label{CoeffSolSDbumpApp2}
\begin{split}
C_0=&-\frac{G_s^2 e^{-\frac{L^2}{\s ^2}} (\lam +\mu ) \left(L^4+4 L^2 \s ^2+8 \s ^4-8 e^{\frac{L^2}{2 \s ^2}} \s ^4+4 e^{\frac{L^2}{\s ^2}} \s ^4\right)}{16 (\lam +2 \mu ) \s ^4}\\
C_1=&\frac{G_s^2 e^{-\frac{L^2}{\s ^2}} \mu  \left(L^4-3 L^2 \s ^2+4 e^{\frac{L^2}{2 \s ^2}} L^2 \s ^2+8 \s ^4-16 e^{\frac{L^2}{2 \s ^2}} \s ^4+8 e^{\frac{L^2}{\s ^2}} \s ^4\right)}{4 L^2 (\lam +2 \mu ) \s ^4}\\
H_1=&\frac{2 G_s  e^{-\frac{L^2}{2 \s ^2}} \left(-\s ^2+e^{\frac{L^2}{2 \s ^2}} \s ^2+L^2 \log{(L)} -2 \s ^2 \log{(L)} +2 e^{\frac{L^2}{2 \s ^2}} \s ^2 \log{(L)} \right)}{L^2 \s ^2}\\
H_2=&\frac{G_s  e^{-\frac{L^2}{2 \s ^2}} \left(L^2-2 \s ^2+2 e^{\frac{L^2}{2 \s ^2}} \s ^2\right)}{L^2 \s ^2}
\end{split}\, .
\eeq
The total energy can be calculated from the general expression \Eqref{Eel} by use of \Eqref{solSDbumpApp} and \Eqref{CoeffSolSDbumpApp1}-\Eqref{CoeffSolSDbumpApp2}, resulting in the two energy profiles plotted in \figref{fig:EnBumpSD}.
\end{widetext}

\bibliography{stick8}
\bibliographystyle{apsrev}

\end{document}